\documentclass[apj]{emulateapj} 

\newcommand{\rmb}{{\rm b}}
\newcommand{\rms}{{\rm s}}
\newcommand{\rmc}{{\rm c}}
\newcommand{\rmd}{{\rm d}}
\newcommand{\rme}{{\rm e}}
\newcommand{\rmm}{{\rm m}}

\newcommand{\beq}{\begin{equation}}
\newcommand{\eeq}{\end{equation}}
\newcommand{\lta}{\la}
\newcommand{\gta}{\ga}
\newcommand{\msunh}{\>h^{-1}{\rm M}_\odot}
\newcommand{\mpch}{\>h^{-1}{\rm Mpc}}

\newcommand{\etal}{{et al. }}
\newcommand{\rmag}{\>^{0.1}{\rm M}_r-5\log h}
\newcommand{\msunhh}{\>h^{-2}{\rm M}_\odot}
\newcommand{\msun}{{\rm M}_\odot}

\usepackage{color}

\shorttitle{Star formation histories}
\shortauthors{Yang et al.}

\begin{document}


\title{Constraining the Star Formation Histories in Dark Matter Halos:
  I. Central Galaxies}

\author{Xiaohu Yang\altaffilmark{1,2}, H.J. Mo\altaffilmark{3}, Frank C. van
  den Bosch\altaffilmark{4}, Ana Bonaca\altaffilmark{4}, Shijie
  Li\altaffilmark{2,6}, Yi Lu\altaffilmark{2}, Yu Lu\altaffilmark{5}, Zhankui
  Lu\altaffilmark{3}}

\altaffiltext{1}{Center for Astronomy and Astrophysics, Shanghai Jiao Tong
  University, 800 Dongchuan Road, Shanghai 200240, China; E-mail:
  xyang@sjtu.edu.cn}

\altaffiltext{2}{Key Laboratory for Research in Galaxies and Cosmology,
  Shanghai Astronomical Observatory, Nandan Road 80, Shanghai 200030, China}

\altaffiltext{3}{Department of Astronomy, University of Massachusetts, Amherst
  MA 01003-9305}

\altaffiltext{4}{Astronomy Department, Yale University, P.O. Box 208101, New
  Haven, CT 06520-8101}

\altaffiltext{5}{Kavli Institute for Particle Astrophysics and Cosmology,
  Stanford, CA 94309, USA}

\altaffiltext{6}{University of Chinese Academy of Sciences, No.19A, Yuquan
  Road, Beijing 100049, China}


\begin{abstract}
  Using the self-consistent modeling of the conditional stellar mass functions
  across cosmic time  by Yang \etal (2012), we make  model predictions for the
  star  formation histories  (SFHs)  of  {\it central}  galaxies  in halos  of
  different masses. The model requires  the following two key ingredients: (i)
  mass assembly  histories of central  and satellite galaxies, and  (ii) local
  observational constraints of the star formation rates of central galaxies as
  function of halo mass. We  obtain a universal fitting formula that describes
  the  (median) SFH  of  central galaxies  as  function of  halo mass,  galaxy
  stellar  mass and  redshift.   We use  this  model to  make predictions  for
  various  aspects of  the star  formation  rates of  central galaxies  across
  cosmic time.   Our main findings are  the following.  (1)  The specific star
  formation rate (SSFR) at high $z$ increases rapidly with increasing redshift
  [$\propto (1+z)^{2.5}$] for halos of a  given mass and only slowly with halo
  mass ($\propto M_h^{0.12}$) at a given $z$, in almost perfect agreement with
  the  specific mass  accretion  rate of  dark  matter halos.   (2) The  ratio
  between the star formation rate  (SFR) in the main-branch progenitor and the
  final stellar  mass of  a galaxy  peaks roughly at  a constant  value, $\sim
  10^{-9.3} h^2 {\rm yr}^{-1}$, independent  of halo mass or the final stellar
  mass of the  galaxy. However, the redshift at which  the SFR peaks increases
  rapidly with halo mass.  (3) More  than half of the stars in the present-day
  Universe were formed in halos with $10^{11.1}\msunh < M_h < 10^{12.3}\msunh$
  in the redshift range $0.4 <  z < 1.9$.  (4) The star formation efficiencies
  (SFE) of central  galaxies reveal a `downsizing' behavior,  in that the halo
  `quenching' mass, at which the SFE peaks, shifts from $\sim 10^{12.5}\msunh$
  at $z \gta 3.5$ to $\sim  10^{11.3} \msunh$ at $z=0$.  (5) At redshift $z\ga
  2.5$ more  than $99\%$ of the  stars in the progenitors  of massive galaxies
  are  formed {\it  in situ},  and this  fraction decreases  as a  function of
  redshift, becoming  $\sim 60\%$  at $z=0$.   For a Milky  Way sized  halo of
  $M_h\sim 10^{12}\msunh$ more than $80\%$ of all the stars are formed {\it in
    situ}, as opposed to having been accreted from satellite galaxies.
\end{abstract}


\keywords{cosmology: dark matter -- galaxies: formation -- galaxies:
  halos}


\section{Introduction}

Recent  years  have seen  dramatic  progress  in  establishing the  connection
between  galaxies and  dark  matter  halos, as  parameterized  either via  the
conditional  luminosity function  (CLF) (e.g.,  Yang \etal  2003) or  the halo
occupation distribution (HOD) (e.g., Jing  \etal 1998; Peacock \& Smith 2000).
This  galaxy-dark  matter connection  describes  how  galaxies with  different
properties occupy halos of  different mass, and contains important information
about how  galaxies form  and evolve  in dark matter  halos. In  practice, the
methods that  have been  used to constrain  the galaxy-dark  matter connection
(galaxy clustering, galaxy-galaxy  lensing, galaxy group catalogues, abundance
matching, satellite kinematics) use the fact that the halo properties, such as
mass  function, mass  profile,  and  clustering, are  well  understood in  the
current $\Lambda$CDM  model of structure formation  (see Mo, \etal  2010 for a
concise review).

At low redshift,  large redshift surveys, such as  the two-degree Field Galaxy
Redshift Survey (2dFGRS; Colless \etal  2001) and the Sloan Digital Sky Survey
(York \etal  2000) have  been used to  establish reliable links  regarding how
galaxies  with different  properties  are distributed  in  halos of  different
masses  (e.g.  Jing \etal  1998; Peacock  \& Smith  2000; Berlind  \& Weinberg
2002;  Yang \etal  2003; van  den Bosch  \etal 2003,  2007; Zheng  \etal 2005;
Tinker \etal 2005;  Mandelbaum \etal 2006; Brown \etal  2008; More \etal 2009,
2011;  Cacciato \etal  2009, 2013;  Neistein et  al.  2011a,b;  Avila-Reese \&
Firmani 2011).  At intermediate redshift, $z\sim 1$, relatively large redshift
surveys,  such as the  DEEP2 survey  (Davis \etal  2003), the  COMBO-17 survey
(Wolf \etal 2004), VVDS (Le Fevre \etal 2005), and zCOSMOS (Lilly \etal 2007),
have  also prompted  a series  of investigations  into the  galaxy-dark matter
connection and its evolution between $z \sim 1$ and the present (e.g., Bullock
\etal  2002; Moustakas \&  Somerville 2002;  Yan \etal  2003; Zheng  2004; Lee
\etal 2006; Hamana \etal 2006; Cooray 2005, 2006; Cooray \& Ouchi 2006; Conroy
\etal 2005, 2007; White \etal 2007; Zheng \etal 2007; Conroy \& Wechsler 2009;
Wang \&  Jing 2010;  Wetzel \&  White 2010; Wake  \etal 2011;  Leauthaud \etal
2012).  However, at higher redshifts, especially beyond $z \simeq 2$, reliable
clustering  measurements  are  not  available,  and the  data  is  limited  to
estimates  of the luminosity/stellar  mass functions  of galaxies  (e.g. Drory
\etal 2005;  Fontana \etal 2006;  Perez-Gonzalez \etal 2008;  Marchesini \etal
2009; Stark  \etal 2009; Bouwens  \etal 2011), often with  large discrepancies
among  different measurements  (see Marchesini  \etal  2009). It  is thus  not
possible to carry out the same  kind of HOD/CLF analyses for high-$z$ galaxies
as for galaxies at low $z$. Nevertheless, attempts have been made to establish
the relation  between galaxies  and their  dark matter halos  out to  high $z$
using a  technique known as abundance  matching (introduced by  Mo \& Fukugita
1996 and Mo, Mao  \& White 1999), in which galaxies of  a given luminosity (or
stellar mass) are linked to dark matter  halos of a given mass by matching the
observed abundance  of the  galaxies to the  halo abundance obtained  from the
halo mass  function (typically  also accounting for  subhalos) (e.g.,  Vale \&
Ostriker 2004, 2006; Conroy \etal 2006; Shankar \etal 2006; Conroy \& Wechsler
2009; Moster \etal 2010; Guo \etal 2010; Behroozi \etal 2010).

However, as pointed out in  Yang \etal (2012; hereafter Y12), although subhalo
abundance matching yields galaxy  correlation functions that are in remarkably
good agreement  with observations  (e.g., Conroy \etal  2006; Guo  \etal 2010;
Wang  \& Jing  2010),  it suffers  from  the following  two problems.   First,
assuming that  the stellar masses of  satellite galaxies depend  only on their
halo mass at accretion implies either that the relation between central galaxy
and halo mass is independent of the time when the subhalo is accreted, or that
the  effects  of  different  accretion  times  and  subsequent  evolutions  in
different hosts conspire to give a  stellar mass that depends only on the mass
of the subhalo at accretion.  Second,  as the satellite galaxies are forced to
be linked with the subhalos that survive in simulations, no satellite galaxies
are allowed  to be associated  with subhalos that  have been disrupted  in the
N-body  simulations.   To circumvent  these  inconsistencies,  Y12 proposed  a
self-consistent model  that properly included  the fact that (1)  the relation
between stellar mass and halo mass of central galaxies depends on $z$; (2) the
properties of  satellite galaxies  depend not  only on the  host halo  mass at
accretion, $m_a$,  but also  on the accretion  redshift, $z_a$; and  (3) after
accretion  a satellite  galaxy  may lose  or  gain stellar  mass  and even  be
disrupted due to  tidal stripping and disruption.  Based on  the host halo and
subhalo accretion  models provided  in Zhao \etal  (2009; see also  Zhao \etal
2003)  and  Yang \etal  (2011),  Y12  obtained  the conditional  stellar  mass
functions  (CSMFs) for  both central  and satellite  galaxies as  functions of
redshift  assuming   a  number   of  popular  $\Lambda$CDM   cosmologies.   In
particular, the mass assembly histories of central galaxies, the population of
accreted satellite galaxies, and  the fraction of surviving satellite galaxies
are all well constrained.

With the  results obtained in  Y12, it is  straightforward to obtain  the star
formation histories  (hereafter SFHs) of galaxies, especially  for the central
galaxies, in halos of different  masses at different redshifts.  Indeed, along
this line, a  couple of recent investigations have tried to  model the SFHs of
galaxies in different  halos using a subhalo abundance  matching (SHAM) method
(e.g., Moster \etal 2013; Behroozi \etal 2012, 2013; Wang \etal 2012).  In all
these  investigations,  assumptions  have  to  be  made  about  the  satellite
evolution, their  contribution to the masses  of central galaxies,  as well as
the intra-halo stars.  In this paper,  we present our own modeling of the SFHs
of (central)  galaxies using a self-consistent  model that is  not hampered by
the shortcomings  of SHAM mentioned  above.  We focus  on the SFHs  of central
galaxies, and defer a discussion of satellite galaxies to a forthcoming paper.

In general  the mass  growth/evolution of a  central galaxy consists  of three
components:  (1) {\it  in situ}  star  formation, (2)  accretion of  satellite
galaxies (`cannibalism'),  and (3)  mass loss due  to the evolution  of stars.
With the  CMSF model presented in Y12,  we can obtain good  constraints on the
growth  of the  central galaxies,  the available  contribution  from satellite
galaxies, as well as the mass loss  due to stellar evolution. With the help of
the observational constraints on the  star formation rates (hereafter SFRs) of
central galaxies in the local Universe, we  will be able to obtain the SFHs of
central galaxies as a function of host halo mass.

This paper is  organized as follows. In Section  \ref{sec_data} we present the
observational  data used  to constrain  the SFRs  of central  galaxies  at low
redshifts.  In Section \ref{sec_CSMF} we describe our methodology to constrain
the  SFRs  of  central  galaxies.    The  results  are  discussed  in  Section
\ref{sec_SFH_data}, where  we also present  an analytical fitting  formula for
the SFRs  of central  galaxies as  function of halo  mass and  redshift. Model
predictions,  including  the  SFHs,  stellar mass  densities,  star  formation
efficiencies, and the fraction of stars formed {\it in situ}, all as functions
of  redshift,   halo  mass  and   stellar  mass,  are  presented   in  Section
\ref{sec_predict}.   Finally, in Section  \ref{sec_discuss}, we  summarize the
main findings of this paper.

Throughout this paper, we use  the $\Lambda$CDM cosmology whose parameters are
consistent  with   the  seventh-year  data   release  of  the   WMAP  mission:
$\Omega_{\rm m}  = 0.275$,  $\Omega_{\rm \Lambda} =  0.725$, $h =  0.702$, and
$\sigma_8 = 0.816$, where the reduced Hubble constant, $h$, is defined through
the Hubble  constant as  $H_0=100h~{\rm km~s^{-1} ~Mpc^{-1}}$  (WMAP7, Komatsu
\etal 2011).  We  use `$\ln$' and `$\log$' to denote  the natural and 10-based
logarithms, respectively.   Unless specified otherwise,  throughout this paper
we use  the following units: SFRs are  in $\msun {\rm yr}^{-1}$,  SSFRs are in
${\rm  yr}^{-1}$,  stellar masses  are  in  $\msunhh$  and have  been  derived
assuming a Kroupa (2001) IMF, and halo masses are in $\msunh$.

\section{Local observational constraints}
\label{sec_data}

\begin{figure*}
\plotone{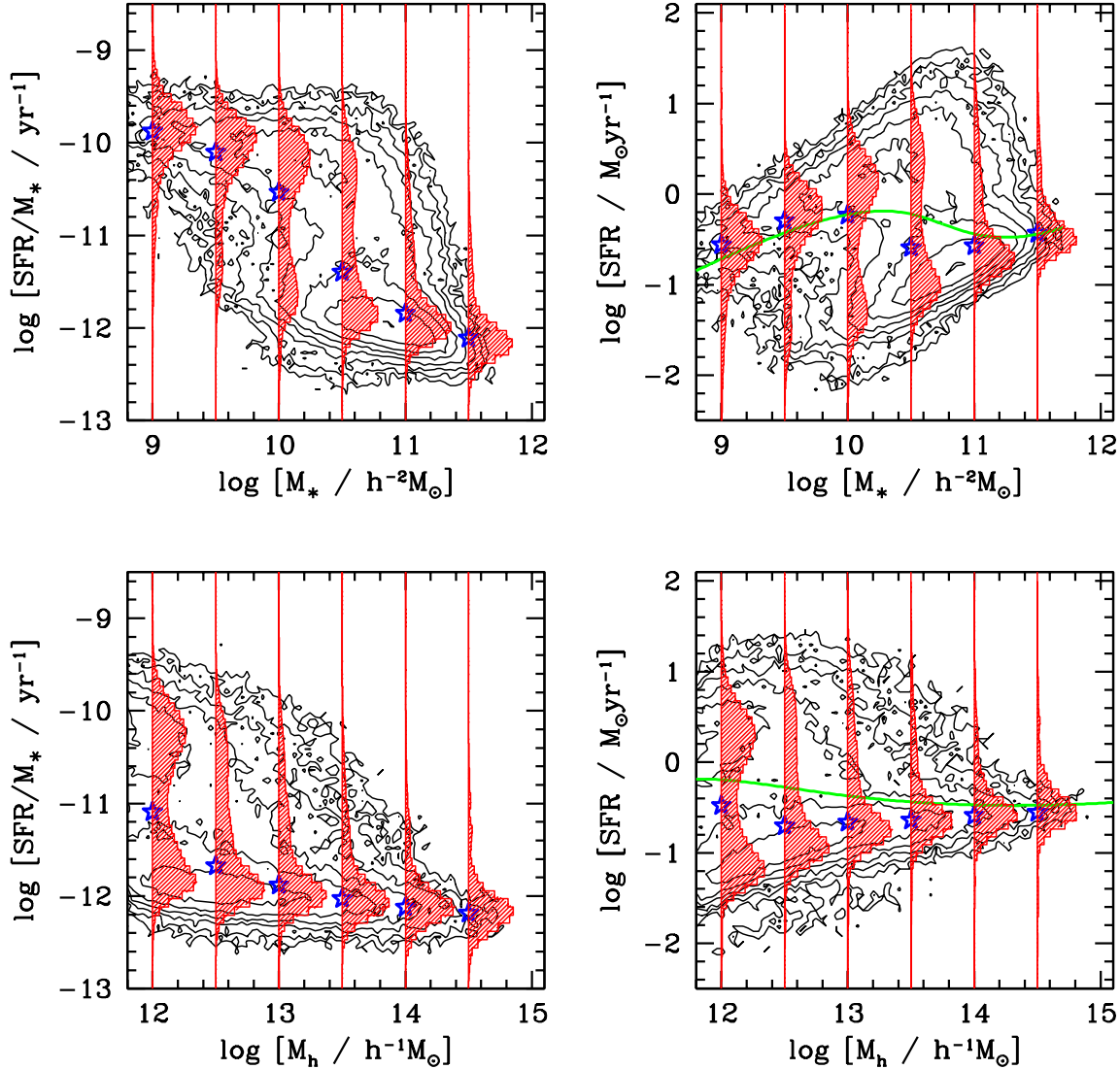}
\caption{The star  formation rates (SFR; right-hand panels)  and specific star
  formation  rates  (SSFR; left-hand  panels)  of  {\it  central} galaxies  as
  function  of their stellar  mass $M_*$  (upper panels)  and halo  mass $M_h$
  (lower  panels).  The  contours  in  each  panel  show  the  number  density
  distribution  of the  galaxies in  logarithmic scale,  with  two neighboring
  levels differing  by a factor two.   The shaded vertical  histograms in each
  panel are the SFR/SSFR distributions  in logarithmic bins of stellar mass or
  halo mass with  widths of $\pm 0.2\,$dex.  The stars  in each panel indicate
  the medians of these distributions,  while the solid lines in the right-hand
  panels show our model prediction (see text). }
\label{fig:SFR_z0}
\end{figure*}

In  order to  model  the SFHs  of central  galaxies  in dark  matter halos  of
different masses, we  need a reliable method to  identify central galaxies, as
well as  accurate estimates of their  SFRs, preferably across  cosmic time. At
low redshift,  such information is  available from galaxy group  catalogs that
can be  used to represent  dark matter  halos.  Here we  make use of  the SDSS
group  catalog constructed  by  Yang  \etal (2007;  hereafter  Y07) using  the
adaptive halo-based group finder developed by Yang \etal (2005).  The original
catalog, based  on SDSS Data Release  4 (DR4) is  updated for DR7\footnote{The
  data  is  available  at {\tt  http://gax.shao.ac.cn/data/Group.html}}.   The
galaxies used  for the  group identifications are  selected from the  New York
University Value-Added  Galaxy catalog (NYU-VAGC; Blanton  \etal 2005b), which
is  based  on  SDSS DR7  (Abazajian  \etal  2009)  but  contains a  number  of
improvements in data  reduction. From the NYU-VAGC, we  select all galaxies in
the  Main  Galaxy  Sample  with  an  extinction-corrected  apparent  magnitude
brighter than $r=17.72$, with redshifts in  the range $0.01 \leq z \leq 0.20$,
and with  a redshift  completeness ${\cal C}_z  > 0.7$.  The  resulting galaxy
catalog contains  a total of $639,359$  galaxies, with a sky  coverage of 7748
square degrees.   A small fraction of  galaxies have redshifts  taken from the
Korea Institute  for Advanced Study  (KIAS) Value-Added Galaxy  Catalog (VAGC)
(e.g.  Choi  \etal 2010).   A total  of $36,759$ galaxies,  which do  not have
redshift  measurements  due  to  fiber  collisions,  are  {\it  assigned}  the
redshifts of their nearest neighbors.  In  the present paper, we use the group
catalog  `modelC', which  is  constructed on  the  basis of  all the  galaxies
(including  those  with  assigned  redshifts)  and uses  model  magnitudes  to
estimate galaxy luminosities.  In total our catalog contains $472,416$ groups,
of which about $23,700$ have three member galaxies or more.  Following Y07, we
assign  a  halo   mass  to  each  group  according  to   the  ranking  of  its
characteristic stellar  mass, defined as the  total stellar mass  of all group
members with $\rmag \leq -19.5$,  where $\rmag$ is the absolute $r$-magnitude,
K- and  E-corrected to $z=0.1$, the  typical redshift of galaxies  in the SDSS
redshift sample.  The  halo mass function adopted in the  ranking is the model
of  Tinker  et  al.   (2008),  which  assumes the  WMAP7  cosmology  and  uses
$\Delta=200$, with $\Delta$ the average  mass density contrast within the halo
(assumed to  be spherically symmetric).  Using the  group catalogue, we
divide galaxies into centrals and satellites; the most massive group member is
identified as the  central galaxy, while all other  group members are assigned
the status of satellite galaxy.  As  already mentioned above, in this paper we
focus on the SFHs of central galaxies only.

The SFRs  and the  specific star  formation rates, SSFRs  (defined as  the SFR
divided by the  stellar mass $M_*$), for individual  galaxies adopted here are
obtained from the data release  of Brinchmann et al.  (2004)\footnote{see {\tt
    http://www.mpa-garching.mpg.de/SDSS/DR7/}},  who  estimated  the  SFR  and
stellar mass  for each  galaxy by  fitting the observed  SDSS spectrum  with a
spectral  synthesis  model. About  94\%  of the  galaxies  used  in our  group
identifications have  estimated SFRs  and SSFRs in  the Brinchmann  \etal data
release  (the vast  majority of  those  lacking estimates  are fiber  collided
galaxies with assigned redshifts).

Fig.  \ref{fig:SFR_z0}  shows the  SSFRs and SFRs  of central  galaxies versus
stellar mass $M_*$ (upper panels) and halo mass $M_h$ (lower panels). In order
to have  a volume-limited sample  of galaxies in  stellar mass, we  adopt, for
given redshift $z$, the following {\it stellar mass} completeness limit, 
\begin{eqnarray} \label{eq:mstarlim}
\lefteqn{\log[M_{*,{\rm lim}}/(\msunhh)] =} \\
 & & {4.852 + 2.246 \log D_L(z) + 1.123 \log(1+z) - 1.186 z \over 1 - 0.067
  z} \, \nonumber
\end{eqnarray}
(see van den Bosch \etal 2008).  In addition, we also apply the following {\it
  halo mass} completeness limit, 
\begin{equation}\label{eq:Mh_limit}
\log [M_{h, {\rm lim}}/(\msunh)] =(z-0.085)/0.069 + 12\,
\end{equation}
(see  Yang  \etal  2009b;  hereafter   Y09b).   The  contours  shown  in  Fig.
\ref{fig:SFR_z0} are  the number density distributions of  central galaxies in
`volume-limited' samples complete in both stellar mass and halo mass according
to $M_{*,{\rm lim}}$ and $M_{h,{\rm lim}}$  given above, i.e.  for a given $z$
we only select systems with $M_\ast > M_{\ast,{\rm lim}}$ and $M_h > M_{h,{\rm
    lim}}$.

As is  evident from  the figure, the  distributions in  both SSFR and  SFR are
bimodal.   For given  $M_\ast$ or  $M_h$, the  central galaxies  appear  to be
separated into  two distinctive populations, one with  high (S)SFRs (hereafter
the ``star forming'' population) and the other with (S)SFRs that are more than
10 times  smaller (hereafter the  ``quenched'' population).  To see  this more
clearly,  we show in  each panel,  using the  vertical shaded  histograms, the
distribution of galaxies within given  logarithmic stellar mass (or halo mass)
bins with widths of $\pm 0.2$dex.   The star plotted on each of the histograms
in each of the panels of  Fig.  \ref{fig:SFR_z0} indicates the median value of
the corresponding distribution.

Let us first focus on the SSFR - stellar mass relation shown in the upper left
panel of  Fig.~\ref{fig:SFR_z0}.  The SSFRs change  significantly with stellar
mass.   Central   galaxies  with  stellar  masses   $\ga  10^{11}\msunhh$  are
predominantly  quenched, while those  with $  M_\ast \la  10^{9.5}\msunhh$ are
mostly star forming. Note, however,  that if satellite galaxies were included,
a significant fraction of the galaxies at the low mass-end would belong to the
quenched population (e.g., van den  Bosch \etal 2008; Peng \etal. 2012; Wetzel
\etal  2012a,b).   Galaxies  with  intermediate  stellar  masses  show  strong
bi-modal  distributions, with  the quenched  population  becoming increasingly
more important as the stellar mass increases.  On average the SSFRs of central
galaxies decrease with increasing stellar mass. Contrary to the SSFR, however,
the   total  SFR   distribution  depicted   in  the   upper  right   panel  of
Fig.~\ref{fig:SFR_z0}  shows  that the  SFRs  increase  roughly linearly  with
stellar mass for the star-forming  population, and with a somewhat slower rate
for the quenched population.

Next, let us  look at how SSFR and  SFR depend on halo mass.  As  shown in the
lower two panels of Fig.  \ref{fig:SFR_z0}, galaxies in halos with masses $\ga
10^{13.0}\msunh$ are dominated by  the quenched population. However, this mode
is much more `stretched' in halo mass than in stellar mass.  This is expected,
because the stellar  mass of central galaxies in  massive halos increases only
slowly with  halo mass,  $M_\ast \propto M_h^{0.22}$,  as shown in  Yang \etal
(2008).

Finally,  we  emphasize once  more  that the  above  results  are for  central
galaxies  only. Had we  included satellite  galaxies, the  quenched population
would have been  significantly more prevalent.  In what  follows these results
will be used  as local observational constraints to model  the SFHs of central
galaxies.

\section{From conditional stellar mass function to star formation histories}
\label{sec_CSMF}

In order to  model the SFHs of  central galaxies as function of  halo mass, we
first  model how  the  stellar components  in  dark matter  halos evolve  with
redshift.   In this  section,  we first  describe  our model  for the  stellar
mass-to-halo mass relation across cosmic  time, based on the results published
in Y12,  followed by  a description  of how such  a model  can be  extended to
describe the SFHs of central galaxies as functions of halo mass.

\subsection{The conditional stellar mass function of galaxies and its evolution}
\label{sec:CSMF}

For  a given  dark matter  halo, the  total stellar  mass it  contains  can be
divided  into  three components:  that  in the  central  galaxy,  that in  the
satellite galaxies, and that in the form of diffuse halo stars. In Y12 we have
developed a  self-consistent model for  the conditional stellar  mass function
(CSMF), which describes the stellar  mass distribution of galaxies in halos of
a  given mass, and  its redshift  evolution.  This  model properly  takes into
account  that (i)  subhalos  are accreted  at  different times,  and (ii)  the
properties of  satellite galaxies may evolve after  accretion. Since satellite
galaxies were  themselves centrals before accretion,  the satellite population
observed today serves as a ``fossil record" for the formation and evolution of
galaxies  in dark  matter  halos over  the  entire cosmic  history. Using  the
observed  galaxy stellar mass  functions out  to $z  \sim 4$,  the conditional
stellar mass function (hereafter CSMF)  at $z\sim 0.1$, obtained from the SDSS
galaxy  group catalogue,  and  the two-point  correlation  function (2PCF)  of
galaxies  at $z  \sim  0.1$ as  function  of stellar  mass,  Y12 obtained  the
relationship between galaxy mass and  halo mass over the entire cosmic history
from  $z \sim  4$  to $z=0.1$.  This relation  was  then used  to predict  the
assembly histories of different stellar mass components (centrals, satellites,
and halo stars) as a function of  halo mass. For completeness, we start with a
brief description of the Y12 results that are relevant for the discussion that
follows.

According  to Y12,  the CSMF  of galaxies  in  halos of  a given  mass can  be
described by the functional form, 
\begin{equation}\label{eq:CSMF_fit}
\Phi(M_{\ast}|M,z) 
= \Phi_\rmc(M_{\ast}|M,z) + \Phi_\rms(m_{\ast}|M,z)\,,
\end{equation}
where   $\Phi_\rmc(M_{\ast}|M,z)$   and   $\Phi_\rms(m_{\ast}|M,z)$  are   the
contributions from the central and satellite galaxies, respectively.  The CSMF
of central galaxies is given by a lognormal distribution, 
\begin{equation}\label{eq:phi_c}
\Phi_\rmc(M_{\ast}|M,z) = {1\over {\sqrt{2\pi}\sigma_c}} {\rm exp}
\left[- { {(\log M_{\ast}/M_{\ast, c} )^2 } \over 2\sigma_c^2} \right]\,,
\end{equation}
where $\log M_{\ast, c}$ is  the expectation value of the (10-based) logarithm
of the  stellar mass of  the central galaxy  and $\sigma_c$ is  the dispersion
(see Y09b).  For simplicity, $\sigma_c$ is  assumed to be  independent of halo
mass, which has observational support  (More \etal 2009).  Following Y09b, the
median stellar mass of the central galaxy is assumed to be a broken power-law,
\begin{equation}\label{eq:Mc_fit}
M_{\ast, c} = M_{\ast, 0} \frac { (M_h/M_1)^{\alpha +\beta} }{(1+M_h/M_1)^\beta } \,,
\end{equation}
so  that  $M_{\ast,  c}  \propto  M_h^{\alpha+\beta}$  ($M_{\ast,  c}  \propto
M_h^{\alpha}$) for  $M_h \ll M_1$ ($M_h  \gg M_1$).  This  model contains four
free  parameters: an  amplitude  $M_{\ast, 0}$,  a  characteristic halo  mass,
$M_1$, and two power indices ,  $\alpha$ and $\beta$.  All four parameters may
depend on redshift, as described below.

The CSMF for satellite galaxies can formally be written as,
\begin{eqnarray} \label{eq:phi_s}
\Phi_{\rm s}(m_{\ast}|M_h,z) =
\int\limits_0^{M_h} \rmd m_a
\int\limits_{z}^{\infty} {\rmd z_a \over 1+z_a} \int\limits_0^{M_h} \rmd M_a
\int\limits_0^1 \rmd\eta ~~~~~~~~~~&&\nonumber\\
\Phi_\rme(m_{\ast}|m_a,z_a,z)  \, n_{\rm sub}(m_a,z_a|M_h,z) ~~~~~~~~~~~~~~~ && \nonumber\\
P(M_a,z_a|M_h,z) \, P(\eta) \, \Theta(p_t \,t_{\rm df} - \Delta t)\,, ~~~~~~~~~~~~~&&
\end{eqnarray}
where $\Theta(x)$  is the Heaviside step  function and $\Delta t$  is the time
between  redshifts   $z_a$  and  $z$.   This  model   contains  the  following
ingredients: (i) the accretion and mass distribution of subhalos, specified by
$n_{\rm  sub}(m_a,z_a|M_h,z) \,  {\rm  d}m_a \,  {\rm  d}\ln(1+z_a)$ which  is
defined as the number  of subhalos in a host halo of  mass $M_h$ identified at
redshift  $z$ as a  function of  their accretion  masses, $m_a$  and accretion
redshifts, $z_a$ (Yang \etal 2011); (ii)  the growth of the main branch of the
host halo,  specified by  $P(M_a,z_a|M_h,z)$, which describes  the probability
that the main  progenitor of a host halo  of mass $M_h$ at redshift  $z$ has a
mass  $M_a$ at  $z_a$ (Zhao  \etal 2009);  (iii) the  orbital  distribution of
accreted  subhalos,  specified  by  $P(\eta)$,  where $\eta$  is  the  orbital
circularity (Zenter \etal 2005); (iv)  the disruption of subhalos during their
evolution  in the  host, which  is specified  by the  dynamical  friction time
$t_{\rm df}$ (Boylan-Kolchin \etal 2008);  (v) the relative disruption rate of
satellite galaxies with respect to  subhalos, which is characterized by a free
parameter $p_t$: $p_t=\infty$ corresponds  to no tidal stripping/disruption of
satellite galaxies,  while $p_t=0$ corresponds to  instantaneous disruption of
all satellites, and finally (vi) the evolution of a satellite from its time of
accretion  until  the  time   corresponding  to  redshift  $z$,  specified  by
$\Phi_\rme(m_{\ast}|m_a,z_a,z)$, the  probability for a subhalo  of mass $m_a$
and  accretion redshift  $z_a$  to host  a  satellite galaxy  of stellar  mass
$m_{\ast}$ at redshift $z$. In Y12 it is assumed that 
\begin{equation}\label{eq:phi_ce}
\Phi_\rme(m_{\ast}'|m_a,z_a,z) = {1\over {\sqrt{2\pi}\sigma_c'}} {\rm exp}
\left[- { {(\log m_{\ast}'/{\overline m}_{\ast}' )^2 } 
\over 2\sigma_c'^2} \right]\,.
\end{equation}
The dispersion, $\sigma_c'$, is assumed to be the same as $\sigma_c$, while
the median is written as
\begin{equation}
{\overline m}_{\ast}' = (1-c) m_{\ast, a} + c m_{\ast, z}\,,
\end{equation}
where $m_{\ast, a}$ is the stellar mass of the satellite at the accretion time
$z_a$, $m_{\ast, z}$  is the stellar mass  of the central galaxy of  a halo of
mass $m_a$  at time $z$,  and $c$  is a free  parameter.  Thus, if  $c=0$ then
${\overline m}_{\ast}' = m_{\ast, a}$ so  that the stellar mass of a satellite
is equal to  its original mass at accretion. Physically  this corresponds to a
picture  in  which  a   satellite  galaxy  is  instantaneously  quenched  upon
accretion. On the other hand,  if $c=1$ then ${\overline m}_{\ast}' = m_{\ast,
  z}$ so that the stellar mass of a satellite is the same as that of a central
galaxy  in a halo  of mass  $m_a$ at  redshift $z$.   This corresponds  to the
assumption made in  SHAM; in other words, by setting  $c=1$ we are effectively
mimicking     SHAM.     Note     that     for    $c=0$     we    have     that
$\Phi_\rme(m_{\ast}'|m_a,z_a,z)$  has the  same form  as the  CSMF  of central
galaxies at the accretion redshift $z_a$.

At any particular redshift, the above  model of the CSMF is fully described by
the following seven free  parameters: $M_{\ast, 0}$, $M_1$, $\alpha$, $\beta$,
$\sigma_c$, $p_t$ and  $c$.  In order to describe the  evolution of the galaxy
distribution over  cosmic time, Y12 assumed the  following redshift dependence
for the model parameters: 
\begin{eqnarray}\label{eq:CSMF_para}
\log[M_{\ast,0} (z)] &=&\log(M_{\ast,0})+ {\gamma_1 z} \nonumber \\
\log[M_1 (z)]        &=&\log(M_1)       + {\gamma_2 z} \nonumber \\
\alpha(z)            &=&\alpha          + {\gamma_3 z} \nonumber \\
\log[\beta(z)] &=&{\rm min} [\log(\beta)+ {\gamma_4 z} + {\gamma_5 z^2} ,2] \\
\sigma_c(z)&=& {\rm max} [0.173, 0.2 z] \nonumber \\
p_t(z) &=& p_t \nonumber \\
c(z) &=& c \nonumber \,.
\end{eqnarray}
The  model is  thus specified  by a  total of  11 free  parameters,  four that
describe the CSMF at $z=0$,  five that describe their evolution with redshift,
and two ($c$ and $p_t$) that describe the evolution of satellite galaxies.

Once the free  parameters are given, the formalism  described above allows one
to  predict the stellar  mass functions  as well  as correlation  functions of
galaxies.  Thus, one can use  observational data on the stellar mass functions
(SMFs)  and  2-point correlation  functions  (2PCFs)  to  constrain the  model
parameters.  Y12  used a Markov Chain  Monte Carlo (hereafter  MCMC) method to
explore  the  likelihood  function  in  the  11-dimensional  parameter  space,
adopting  the WMAP7  cosmology.  In  particular,  two types  of analysis  were
carried out  for comparison.  The first  uses the SMFs  at different redshifts
together with the 2PCFs of galaxies at low-$z$ as constraints; the second uses
the SMFs at different redshifts together with the CSMFs at low $z$. Since both
analyses gave  very similar  results, in  this paper we  only use  the results
based on the latter analysis.  Furthermore, Y12 used two sets of high redshift
stellar mass  functions, one obtained by Perez-Gonzalez  \etal (2008; referred
to  as SMF1), and  the other  obtained by  Drory \etal  (2005; referred  to as
SMF2). Since these two data sets reveal quite large discrepancies with respect
to  each other  and  dominate over  all  the uncertainties  in our  model
  ingredients,  we will present results separately for both of them.

\subsection{The growth of stellar components in dark matter halos}
\label{sec:growth}

Once the redshift-dependent CSMFs are  obtained, one can predict the growth of
the  stellar masses  of both  central and  satellite galaxies  along  the main
branch of their dark matter halos.  The median stellar mass at redshift $z$ of
a central galaxy  that at redshift $z_0 \leq  z$ is located in a  host halo of
mass $M_0$ can be written as 
\begin{equation}\label{cengrowth}
M_{\ast, c} (z|M_0,z_0) = M_{\ast, c} (M_a,z)\,.
\end{equation}
Here, as before, $M_a$ is the median mass at redshift $z \geq z_0$ of the main
progenitor of  a host halo of mass  $M_0$ at redshift $z_0$.   The median {\it
  total} stellar mass  of the surviving satellite galaxies  in the main branch
can be obtained by integrating the CSMF of satellite galaxies: 
\begin{equation}\label{satgrowth}
 m_{\ast, s} (z|M_0,z_0) = \int \rmd \log m_{\ast} \, m_{\ast} \,
 \Phi_\rms(m_{\ast}|M_a,z) \,.
\end{equation}
Thus,  once the  assembly  history  of a  dark  matter halo  is  known, it  is
straightforward to use the CSMF to obtain the corresponding assembly histories
of the stellar  components. In addition, one can also  estimate the total mass
of all satellite galaxies that have  been accreted into the main branch, which
includes stellar  mass in  both the currently  surviving satellites  and those
that have  been cannibalized by the  central galaxy or disrupted  by the tidal
field. This is given by 
\begin{eqnarray}\label{satacc}
m_{\ast, {\rm acc}} (z|M_0,z_0) = \int_{z_0}^{z}
{\rmd z_a \over (1+z_a)} \int \rmd \log m_{\ast} \, m_{\ast} ~~~~~~~~~~&&\\
\Phi_\rme(m_{\ast}|m_a,z_a,z)  \,
n_{\rm sub}(m_a,z_a|M_0,z_0)\,.~~~~~~~~~~~~~~\nonumber
&&
\end{eqnarray}
Note  that  $\Phi_\rme$  accounts  for  evolution in  the  stellar  masses  of
satellite  galaxies  after  accretion  due  to star  formation,  stellar  mass
stripping and mass loss due to stellar evolution.

The difference between $m_{\ast, {\rm acc}}$ and $m_{\ast, s}$ gives the total
mass of (destroyed) satellites that  are either cannibalized by the central or
disrupted  by the tidal  field.  As  shown in  Y12, the  mass in  this stellar
component in  a massive cluster  can be much  larger than that of  the central
galaxy.   Hence, a significant  fraction of  the total  stellar mass  from the
disrupted satellite galaxies cannot be associated with the central galaxy, but
instead has  to be  in the  form of diffuse  halo stars.   Unfortunately, what
fraction of stars  in a halo is associated with such  a diffuse halo component
(also called  `intra-cluster light' in the  case of clusters)  is still poorly
constrained observationally (e.g.  Gonzalez  \etal 2005 Seigar \etal 2007), so
that  it  remains  unclear  what  fraction of  the  `destroyed'  satellite  is
cannibalized by the central vs. added  to the stellar halo (e.g. Purcell \etal
2007; Kang \&  van den Bosch 2008; Yang \etal  2009a). Two extreme assumptions
can be made for the contribution  of the destroyed satellites to the growth of
the central:  (i) minimum (zero) contribution; and  (ii) maximum contribution,
where the contributed  mass is equal to either the mass  growth of the central
or the  total mass of destroyed  satellites, whichever is smaller,  in a given
time interval.

In general,  the growth  (evolution) of the  central galaxies consists  of the
following three contributions: (i) its  {\it in situ} star formation; (ii) the
accretion of  stars from satellite  galaxies; and (iii) its  passive evolution
(mass loss). The model described above for the stellar mass assembly histories
of central galaxies and the possible contribution from accreted satellites can
therefore be used to infer the SFH of central galaxies: 
\begin{eqnarray}\label{eq:SFH_model}
 SFR(z) &=& \frac{\rmd M_{\ast,c}(z)}{\rmd t} -f_{\rm acc} \left\{ 
\frac{\rmd [m_{\ast,{\rm acc}}(z) -  m_{\ast,s}(z)]}{\rmd t} \right\} \nonumber \\ 
&+&  \frac{\rmd M_{\ast, {\rm loss}}(z)}{\rmd t} \,,
\end{eqnarray}
where
\begin{equation}\label{eq:mass_loss}
 \frac{\rmd M_{\ast, {\rm loss}}(z)}{\rmd t} = \int^{t(z)}_0
 \frac{\rmd M_{\ast,c}(z)}{\rmd t_1} \,
 \frac{\rmd f_{\rm passive}(t(z)-t_1)}{\rmd t} \rmd t_1\,,
\end{equation}
is the stellar mass loss due to the passive evolution of stars.  Corresponding
to the two extreme cases for  the contribution of the destroyed satellites are
two  extreme estimates  for the  SFH:  (i) maximum  star formation  (hereafter
`MAX'), in which  the stellar mass growth of central  galaxies is entirely due
to  {\it in situ}  star formation,  i.e.  $f_{\rm  acc}=0$; (ii)  minimum star
formation  (hereafter   `MIN'),  in  which  the   contribution  from  accreted
satellites  is maximized ($f_{\rm  acc}=1$) or  $SFR(z)=0$ if  $f_{\rm acc}=1$
leads to $SFR(z)<0$.

The  stellar mass  loss due  to  passive evolution  is accounted  for via  the
function $f_{\rm passive}(t)$, which describes the mass fraction of stars that
at time  $t$ after their formation  is still in  the form of stars.  We obtain
$f_{\rm passive}(t)$ from  the stellar population model of  Bruzual \& Charlot
(2003),     kindly      provided     by     Stephane      Charlot     (private
communication).  Throughout this  paper we  adopt the  results for  the Kroupa
(2001) IMF, which asymptotes to $f_{\rm  passive} \sim 0.57$ at late stages of
evolution.  For  Chabrier (2003) and Salpeter (1955)  IMFs, these asymptotical
values  are $\sim 0.54$  and $\sim  0.70$, respectively.  We have  tested that
changing the IMF from Kroupa to Chabrier or Salpeter results in changes in the
SFRs of $\la  0.1\,{\rm dex}$, much smaller than  the uncertainties from other
sources, e.g., the use of SMF1 or SMF2.

\section{The  Star Formation Histories of Central Galaxies}   
\label{sec_SFH_data}

\begin{figure*}
\plotone{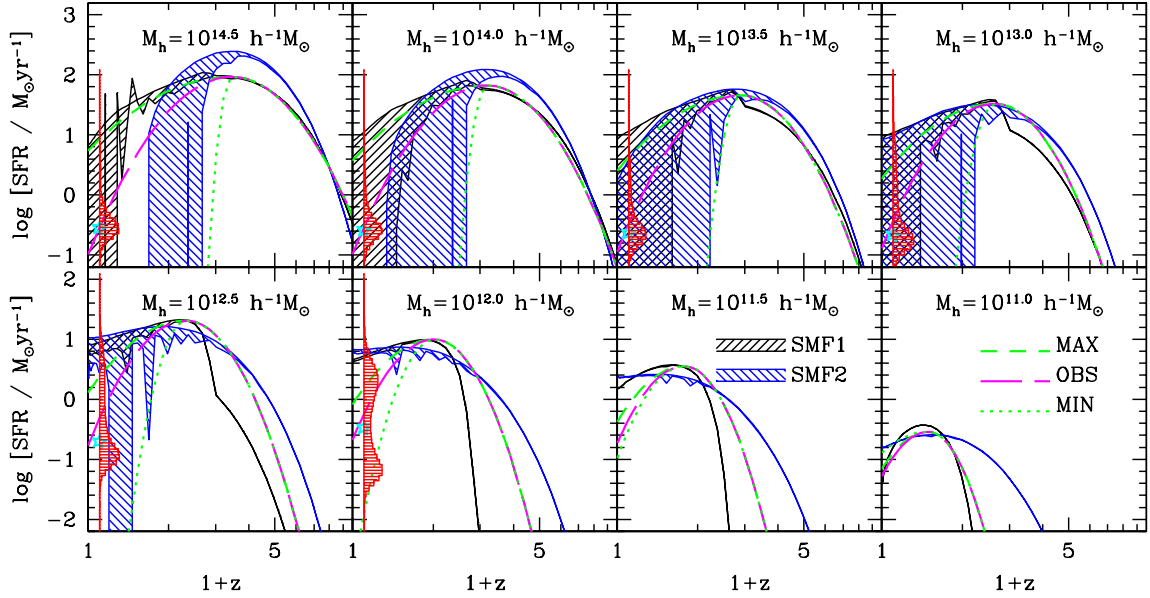}
\caption{The star formation rates (SFRs)  of central galaxies as a function of
  redshift  in halos  of different  present-day masses,  as indicated  in each
  panel. Black and blue shaded  regions reflect the model predictions obtained
  using SMF1 and  SMF2, respectively. For each of these  two cases, the shaded
  areas  mark the  SFRs between  the  MIN and  MAX assumptions  (see text  for
  details),  which become more  and more  similar at  higher redshift  and for
  lower mass  halos. In  the panels for  halos with $M_h  \geq 10^{12}\msunh$,
  local observational constraints  from SDSS are shown as  the vertical shaded
  histograms  (distributions) and  stars  (median).  The  dotted, dashed,  and
  long-dashed lines are the MIN, MAX and OBS fits to the SFHs discussed in the
  text.}
\label{fig:SFH}
\end{figure*}
\begin{figure*}
\plotone{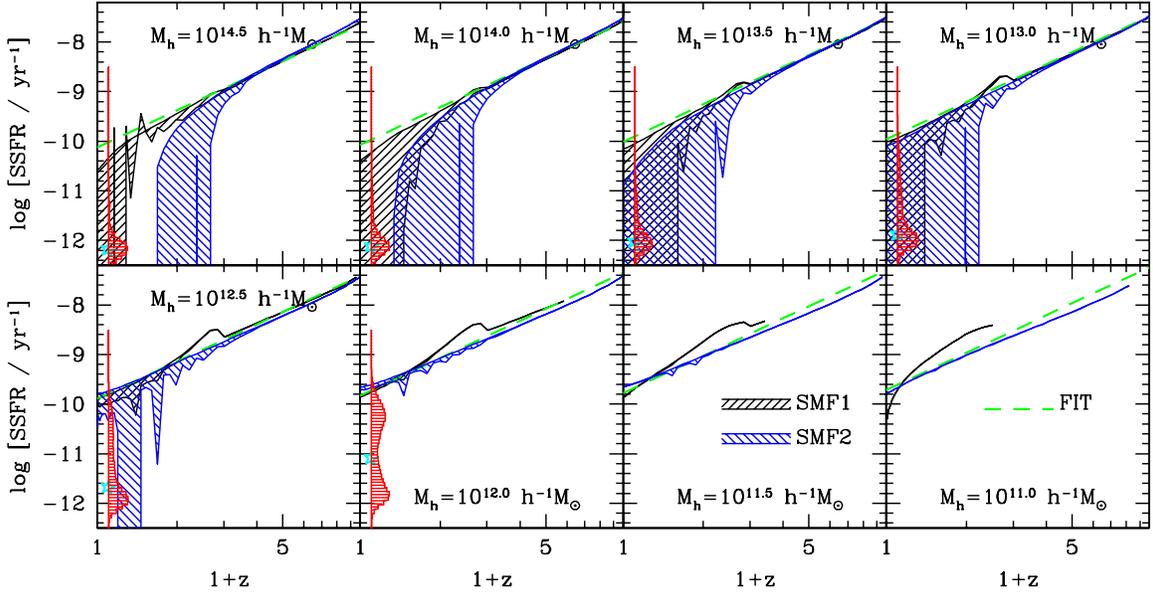}
\caption{Same  as Fig,~\ref{fig:SFH}, except  that here  we show  the specific
  star formation  rates (SSFRs) of  central galaxies as function  of redshift.
  The  dashed  line  is  the  model  fit  of  Eq.~(\ref{equ:SSFR_fit}),  which
  accurately describes the SSFRs of  central galaxies prior to their quenching
  (see text for detailed discussion).}
\label{fig:SSFH}
\end{figure*}
\begin{figure*}
\plotone{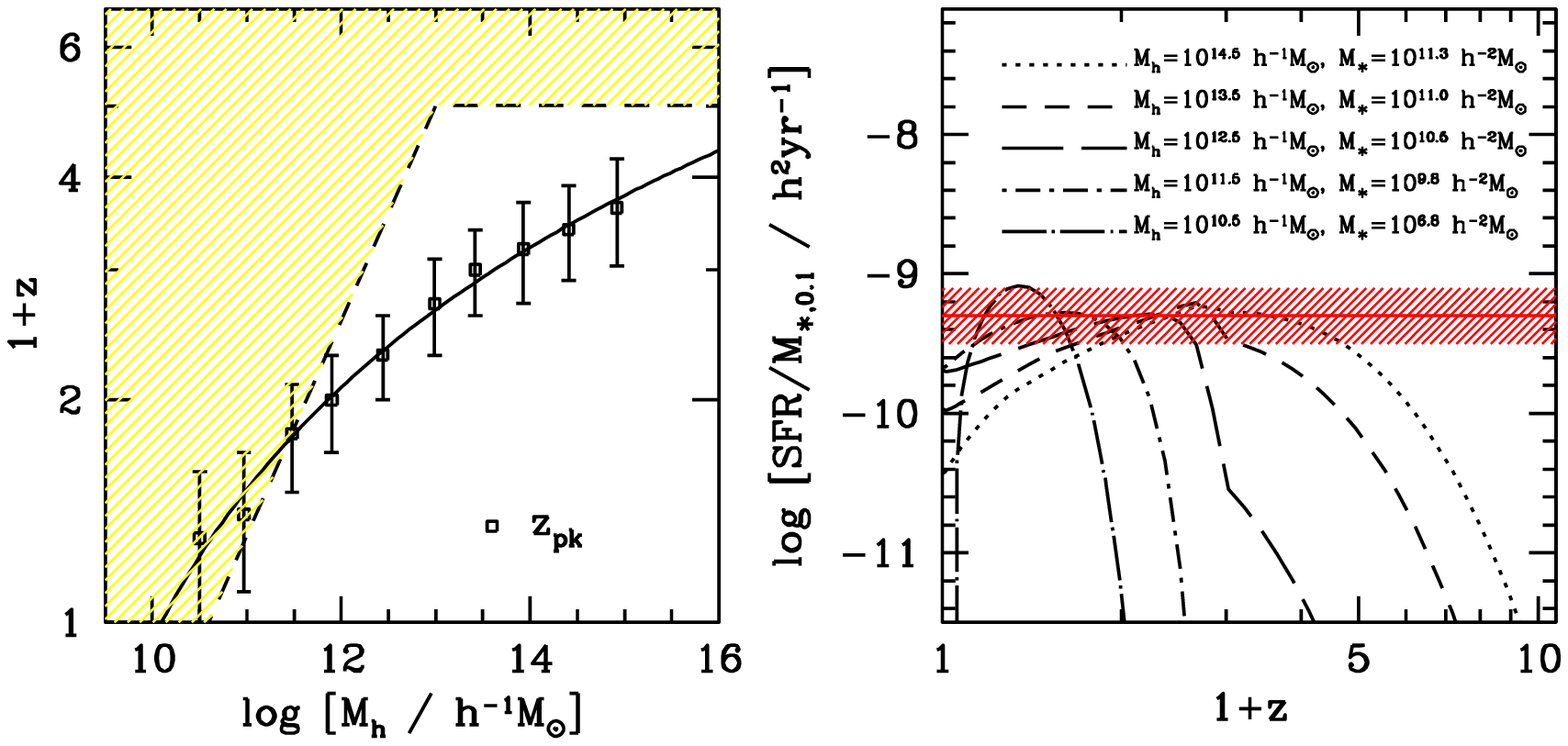}
\caption{{\it Left-hand panel:} the peak redshift  of the SFR as a function of
  present-day  host halo  mass.  Data  points with  error bars  are  the rough
  estimations of the peak redshifts from the SFHs shown in Fig.~\ref{fig:SFH},
  with the  errorbars reflecting the redshift  interval over which  the SFR is
  within 10\% of its  peak value.  The solid line is a  fit to the data points
  (Eq.~[\ref{equ:SFR_zpeak}]).   As an illustration,  we also  show using
    the   shaded   region  where   our   SFH   results   are  obtained   from
    extrapolations.   {\it Right-hand panel:} the star formation histories of
  central galaxies, normalized  by the stellar masses of  the central galaxies
  at redshift  $z=0.1$.  Over  about 4  orders of magnitude  in halo  mass and
  stellar  mass,  these  normalized  SFHs peak  at  $10^{-9.3}  h^{2}{\rm
      yr}^{-1}$  (within $\sim  0.2\,$dex),  as indicated  by the  horizontal
  band.}
\label{fig:SFH_peak}
\end{figure*}

Applying  the model  outlined  in Section~\ref{sec:growth}  from  high to  low
redshifts,  we obtain  the  SFHs of  central  galaxies in  halos of  different
masses.   Figs.~\ref{fig:SFH} and  ~\ref{fig:SSFH}  show the  median SFRs  and
SSFRs as a  function of redshift for the centrals in  halos of different final
masses, as indicated  in each panel.  In each panel, the  black and blue lines
correspond to the results obtained using SMF1 and SMF2, respectively. For each
set of  observational data (SMF1 or  SMF2), predictions based on  both the MIN
and MAX  assumptions are shown,  connected by the corresponding  shaded areas.
As  described  in Section~\ref{sec:growth},  MIN  and  MAX  correspond to  the
maximum and  minimum (zero) contribution of  satellites to the  mass growth of
central galaxies,  respectively, so  that they represent  the lower  and upper
limits on the SFRs of central galaxies.  As one can see, at $z>2$, these upper
and  lower limits  are always  very  similar for  halos of  all masses.   This
indicates  that  the  growth  of  stellar  mass at  these  high  redshifts  is
completely  dominated by {\it  in situ}  star formation  (see Y12).   At lower
redshift,  the MIN  and  MAX  assumptions results  in  fairly different  SFRs,
especially  for halos  with $M_h  \gta 10^{12}\msunh$.   In these  halos, mass
growth due to  the accretion of satellite galaxies could  be responsible for a
significant fraction of the final stellar mass of centrals.  In the absence of
accurate estimates for the mass components in halo stars, it will be difficult
to discriminate between these MIN and MAX models, and therefore to tighten the
constraints.

However, as  shown in Section~\ref{sec_data},  the (S)SFRs of  low-$z$ central
galaxies  can  be  estimated  for  halos of  different  masses  directly  from
observational data.   This provides a direct  constraint on the  SFH at $z\sim
0$.  In Figs.~\ref{fig:SFH} and ~\ref{fig:SSFH} the vertical shaded histograms
show  the   local  observational  constraints  for  halos   with  masses  $\ga
10^{12}\msunh$.  Results  for halos with  lower masses are not  available from
the group catalog (see Y07).  In general, the median (S)SFRs obtained directly
from the  SDSS group  catalog (indicated by  crosses) nicely falls  within the
shaded area, between  the upper and lower limits corresponding  to the MIN and
MAX  assumptions,  indicating  that  the  SFRs obtained  from  our  model  are
consistent with the results obtained  directly from the SDSS groups.  However,
there  are two  exceptions:  our  model under-predicts  the  median (S)SFR  of
centrals in massive halos with $M_h \ga 10^{14.0} \msunh$ in the case of SMF2,
and over-predicts the median (S)SFRs of  central galaxies in halos with $M_h =
10^{12}\msunh$.  In the  case of SMF2, the model prediction  of the (S)SFR for
central  galaxies in massive  halos with  $M_h \ga  10^{14.0} \msunh$  is well
below  the median  (S)SFRs  obtained  from the  group  catalog.  This  happens
because  the use of  SMF2 slightly  over-predicts the  number density  of very
massive galaxies at redshift $z\sim 1$ in comparison to that obtained directly
from the SDSS data.

For halos with $M_h \sim 10^{12}\msunh$ the models based on both SMF1 and SMF2
predict  a  median SSFR  at  $z=0.1$ of  $\sim  10^{-10}  {\rm yr}^{-1}$.   In
contrast,  the  SSFRs  obtained  directly  from  the  SDSS  data  reveals  two
distinctive populations: a star-forming  population, whose SSFRs peak at $\sim
10^{-10.2} \,{\rm yr}^{-1}$ and a quenched population with $\langle {\rm SSFR}
\rangle  \sim 10^{-11.8}\,{\rm  yr}^{-1}$.  Thus,  it appears  that  our model
fails to account for the quenched  population.  We suspect that this is caused
by two effects.  First, due to  contamination in the group finder, some of the
red centrals in  the group catalog are actually  satellite galaxies.  However,
as shown in Y07, this contamination fraction is about $\la 10\%$, which is not
sufficient to  explain the relatively  large fraction of quenched  centrals in
groups with $M_h  \sim 10^{12}\msunh$.  Second, some of  the quenched centrals
may be  a population of galaxies  that is located  in the outskirts of  a more
massive halos which it traversed in  the past. Star formation in such galaxies
is likely  to have experienced the  same quenching as  satellites during their
journeys through the massive halos (Wetzel \etal 2013).  Numerical simulations
suggest that  such a  population is  expected, as some  of the  low-mass halos
located near  massive ones have indeed  passed through massive  halos at least
once in the past (e.g.  Lin \etal 2003; Gill, Knebe \& Gibson 2005; Wang \etal
2009a;  Ludlow \etal 2009).   Observationally, Wang  \etal (2009b)  found that
many  of the  red dwarf  galaxies  not contained  within the  virial radii  of
massive halos are located within three virial radii from nearby massive halos,
consistent  with the idea  that they  were quenched  when they  passed through
their massive neighbor (see also Geha  \etal 2012).  If such a population also
exist  for galaxies  with stellar  masses $\sim  10^{10}\msunhh$, it  might be
possible to explain the quenched  population of centrals in galaxy groups with
masses $\sim 10^{12}\msunh$.  Such a  population is missed in our model, where
all halos  that have  once been accreted  into a  host halo are  considered as
subhalos hosting satellite galaxies.   Clearly, detailed analysis is needed in
order to quantify the contribution of this kind of ``passing-through" galaxies
to the quenched population (see e.g., Wetzel \etal 2013).

\subsection{Analytical Model}
\label{sec_model}

Let us now move on to a  quantification of the predicted SFHs.  We first focus
on the SSFRs of central galaxies. As one can see from Fig.  \ref{fig:SSFH}, at
high $z$  the SSFRs roughly  reveal a power  law dependence on  redshift.  The
results  obtained from  SMF1 and  SMF2 are  remarkably similar,  and  are well
described by 
\begin{eqnarray}\label{equ:SSFR_fit}
\log ({\rm SSFR}/{\rm yr}^{-1}) &=& 2.5\log (1+z) \\
  &-& 0.12\log (M_h/\msunh) -12.0\,, ~~~~~\nonumber
\end{eqnarray}
which is indicated as the dashed line  in each of the panels in the figure. In
small halos with $M_h \la 10^{12.0}\msunh$, Eq.~(\ref{equ:SSFR_fit}) holds for
the  entire   SFH.   In  halos   with  $M_h  \ga   10^{12.5}\msunh$,  however,
Eq~(\ref{equ:SSFR_fit}) holds only for $z > z_{\rm pk}$, where $z_{\rm pk}$ is
defined    as    the    redshift    at    which    the    SFR    peaks    (see
below). Eq.~(\ref{equ:SSFR_fit})  indicates that the SSFR  of central galaxies
depends only  weakly on galaxy mass.   For example, central  galaxies in halos
with $M_h = 10^{11}$ and $10^{12}\msunh$ have similar SSFRs, even though their
stellar masses differ by more than an order of magnitude (cf. Fig.~13 in Y12).
The dependence of SSFR on redshift is much stronger, increasing by a factor of
about 6  from $z=1$ to  $z=3$.  Interestingly, this  scaling of the  SSFR with
halo  mass and  redshift is  almost  identical to  that of  the specific  mass
accretion rate  of dark matter  halos, which scales as  $\dot{M}_h/M_h \propto
M_h^{0.15} \, (1+z)^{2.25}$ (Dekel \etal  2009; see also McBride \etal 2009; 
Fakhouri \etal 2010 for similar results obtained from N-body simulations). 
This suggests that the SFR of
star-forming (i.e., non-quenched) central galaxies is regulated by the rate at
which their host  halos accrete mass, in excellent agreement  with a number of
recent  studies (e.g., Dutton,  van den  Bosch \&  Dekel 2010;  Bouch\'e \etal
2010; Dav\'e, Finlator \& Oppenheimer 2012).

The  SFRs  in  Fig.~\ref{fig:SFH}   show  that  $z_{\rm  pk}$  increases  with
increasing stellar mass (or halo  mass), indicating that more massive centrals
are quenched earlier  (a manifestation of what is  often called `downsizing').
We have  estimated $z_{\rm pk}$ as a  function of halo mass,  $M_h$, using the
SFHs obtained  from both  SMF1 and  SMF2.  The open  squares in  the left-hand
panel of Fig.~\ref{fig:SFH_peak} show the average between SMF1 and SMF2, while
the errorbars are obtained from the points  where the SFH is about 10\% of the
peak value.  Using the following functional form to fit the data, 
\begin{equation}\label{equ:SFR_zpeak}
z_{\rm pk} = \max [ a (\log M_h - b), 0] \,,
\end{equation}
we  obtain $a=0.568$ and  $b=10.10$.  This  fitting function  is shown  in the
left-hand panel of Fig.~\ref{fig:SFH_peak} as the solid line.

In addition to  $z_{\rm pk}$ increasing with halo mass, the  actual SFR at the
peak  redshift   also  increases   with  halo  mass.    As  is   evident  from
Fig.~\ref{fig:SFH}, this  halo mass dependence becomes weaker  in more massive
halos. This kind of halo mass dependence is also seen for the luminosities and
stellar  masses  of  central  galaxies  at low  redshift  (see,  e.g.,  Y09b).
Interestingly,  if we  normalize the  SFRs of  the central  galaxies  by their
stellar masses at $z=0.1$, the peak amplitude becomes virtually independent of
halo mass and  stellar mass!  This is demonstrated in  the right-hand panel of
Fig.~\ref{fig:SFH_peak}, which plots $\log[SFR/M_{{\ast},0.1}]$ as function of
redshift for five different values of the present-day halo mass, as indicated.
The  peak  amplitude  is  about   $\log[{\rm  SFR}/M_{{\ast},  0.1}\times
    h^{-2}{\rm yr} ]= -9.3$, which  is indicated by the horizontal line.  The
hatched, horizontal  band covers  $\pm 0.2$ dex  around this line  and roughly
represents the uncertainty in the estimate of the SFR and stellar mass.  Thus,
central galaxies with stellar mass  $M_{\ast, 0.1}$ at $z=0.1$, have a maximum
SFR 
\begin{equation}\label{equ:SFR_amp}
{{\rm SFR}_{\rm pk}\over [\msun\,{\rm yr}^{-1}] }
= {M_{\ast, 0.1}\over 10^{9.3}\msunhh}\,
\end{equation}
at redshift  $z_{\rm pk}$ given by  Eq.~(\ref{equ:SFR_zpeak}). These equations
hold for  central galaxies covering  over 4 orders  of magnitude in  both halo
mass and stellar mass.  Note that our modeling here is based mainly on results
obtained from SMF1.  The results obtained from SMF2 are qualitatively similar,
and shown in the Appendix for comparison.

Motivated  by  the  general appearance  of  the  SFHs,  we use  the  following
functional form to model the SFHs of central galaxies as function of halo mass
and redshift: 
\begin{equation}\label{equ:SFH}
{\rm SFR}(M_h, z) = {\rm SFR}_{\rm pk} 
\times\exp\left\{-{\log^2[(1+z)/(1+z_{\rm pk})]
\over 2\sigma^2(z_{\rm pk})}\right\}\,,
\end{equation}
where $\sigma(z_{\rm pk})$ describes the decay  of the SFR with respect to the
peak. By trial-and-error,  we find that a simple power-law  scaling with $(1 +
z_{\rm  pk})$ can  adequately describe  the SFHs.   For $z\ge  z_{\rm  pk}$ we
obtain 
\begin{equation}\label{equ:decay1}
\sigma (z_{\rm pk})= 0.0576 (1+z_{\rm pk})^{0.707}\,,
\end{equation}
where the parameters  are obtained using least squares fitting  to the SFHs at
high redshifts.  The  differences between the SMF1 and  SMF2 model predictions
are used as  weights in the fitting.  For $z< z_{\rm  pk}$ we obtain different
power-law scalings: 
\begin{eqnarray}\label{equ:decay}
\sigma (z_{\rm pk})= \left\{\begin{array}{ll} 
 0.0762 (1+z_{\rm pk})^{0.523}~~~ &\mbox{(OBS)}\\
  0.0706 (1+z_{\rm pk})^{0.940}~~~ &\mbox{(MAX)}\\
 0.317 (1+z_{\rm pk})^{-2.10}~~~ &\mbox{(MIN)}
\end{array}\right.\,.
\end{eqnarray}
Here the case marked `OBS' assumes that the SFHs at low-$z$ are constrained by
the SFR measurements  from the SDSS groups using least  squares fitting to the
median SFRs  of central galaxies as  a function of stellar  mass (asterisks in
upper  right-hand  panel of  Fig.~\ref{fig:SFR_z0}).   The  case marked  `MAX'
assumes that all stars in central  galaxies are formed {\it in situ}, in which
case  the stellar mass  of central  galaxies as  function of  redshift, $z_0$,
follows from  integrating the SFH  and taking into  account the effect  due to
passive evolution, 
\begin{equation}\label{equ:M_*01}
M_{\ast}(z_0) = \int_0^{t_{z_0}} {\rm SFR}_{\rm MAX}(t) \, 
f_{\rm passive}(t_{z_0}-t)~{\rm d}t \,.
\end{equation}  
Here $t_{z_0}$ is  the age of the  universe at redshift $z_0$ and  $1 - f_{\rm
  passive}(\Delta t)$ is the mass fraction of stars formed that a time $\Delta
t$ later has  been returned to the IGM due to  stellar (passive) evolution. By
definition, $M_{\ast}(z_0=0.1) =  M_{\ast, 0.1}$, which is used  to obtain the
two parameters in Eq.(\ref{equ:decay}) for the `MAX' case. For the `MIN' case,
the   two  parameters   in  Eq.\,(\ref{equ:decay})   are  obtained   from  the
corresponding low-redshift SFHs given by SMF1.

The  resulting SFH  model predictions  are plotted  as the  long-dashed (OBS),
dashed  (MAX)   and  dotted  lines   (MIN)  in  Fig.    \ref{fig:SFH}.   Thus,
Eq.~(\ref{equ:SFH}) describes the {\it median} SFH for central galaxies with a
given stellar mass at redshift $z=0.1$.   For halos of a given mass $M_h$, one
can    first    obtain     the    corresponding    $M_{\ast,    0.1}$    using
Eqs.\,(\ref{eq:Mc_fit}) and~(\ref{eq:CSMF_para}), and then obtain the SFH from
Eq.\,(\ref{equ:SFH}).  The uncertainties  in the SFH model may  be gauged from
the differences between  the model predictions for SMF1  and SMF2, and between
the   results  for   the   two   extreme  assumptions,   MIN   and  MAX   (see
Fig.~\ref{fig:SFH}). As pointed  out in Y12, at the  present the uncertainties
in the CSMF modeling are dominated  by systematic errors between SMF1 and SMF2
(see Fig.~14 in Y12), rather than by the statistical errors in the data.  Note
that our  model is  obtained from observational  measurements of  stellar mass
functions that are limited in both redshift  ($z \la 4$ for SMF1 and $z \la 5$
for SMF2) and stellar mass (see  Fig.3 in Y12).  Results beyond these redshift
and/or  stellar mass  ranges are  in  general obtained  by extrapolation,  and
therefore less reliable.  As an  illustration, we show, in the left panel
  of Fig.  \ref{fig:SFH_peak}  using the shaded region, where  our SFH results
  are obtained from extrapolations. 

Finally, we  emphasize that our  model for the  SFHs is based on  the assembly
histories  of central  galaxies in  halos of  a given  final mass,  and  it is
expected to work well only for  centrals with mass assembly histories close to
the median. Deviations from the median are expected to result in deviations of
the inferred SFHs from the model prediction and may in fact be the main source
of the scatter seen in Fig. \ref{fig:SFR_z0}.  In a forthcoming paper, we will
come back to the modeling of  this scatter together with the SFHs of satellite
galaxies.

\begin{figure*}
\plotone{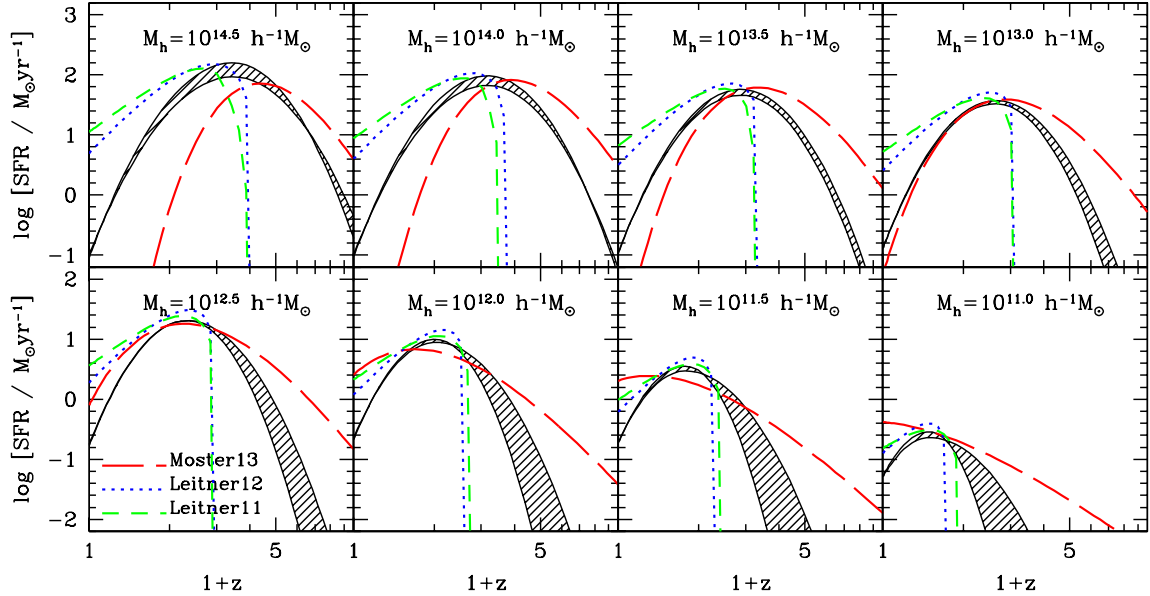}
\caption{The SFHs of central  galaxies, similar to Fig.~\ref{fig:SFH} but here
  our model predictions are compared with the results obtained in a few recent
  papers. The lower and upper boundaries  of the shade areas are our OBS model
  predictions of  the SFHs  for central  galaxies based on  SMF1 with  the SFR
  described  by  Eq.~(\ref{equ:SFH}),  and  SMF2  with the  SFR  described  by
  Eq.~(\ref{equ:SFH2}),  respectively.   The dashed,  dotted  and long  dashed
  lines  in each  panel are  model predictions  based on  Leitner  \& Kravtsov
  (2011), Leitner  (2012) and Moster  et al.  (2013),  respectively. See
  text for a detailed discussion.}
\label{fig:SFH_comp}
\end{figure*}

\subsection{How to Use the Model}
\label{sec:use}

As  a summary,  the following  is  a brief  description of  the procedure  for
obtaining the median SFH for a central galaxy in a halo of a given mass:

\begin{enumerate}

\item Start from a  halo with mass $M_h$ at low redshift  (e.g. $z \sim 0.1$).
  Use Eqs. (17) and~(40) in Yang \etal (2012), which are the relations adopted
  in this  paper, or use Eq.\, (20)  of Yang \etal (2009),  which was obtained
  directly from the  SDSS group catalog, or use any other  stellar mass - halo
  mass relation  for central galaxies,  to obtain the stellar  mass, $M_{\ast,
    0.1}$, of the central galaxy at $z \simeq 0.1$;

\item Use  Eq.~(\ref{equ:SFR_zpeak}) to obtain the redshift,  $z_{\rm pk}$, at
  which the SFR peaks;

\item  Use Eq.~(\ref{equ:SFR_amp}) to  get the  peak value  of the  SFR, ${\rm
  SFR}_{\rm pk}$;

\item  Finally,  the  median  SFH  of  the  central  galaxy  is  described  by
  Eq.~(\ref{equ:SFH})     with    $\sigma     (z_{\rm    pk})$     given    by
  Eqs.~(\ref{equ:decay1}) and~(\ref{equ:decay}) for the case labelled `OBS'.

\end{enumerate}

\begin{figure*}
\plotone{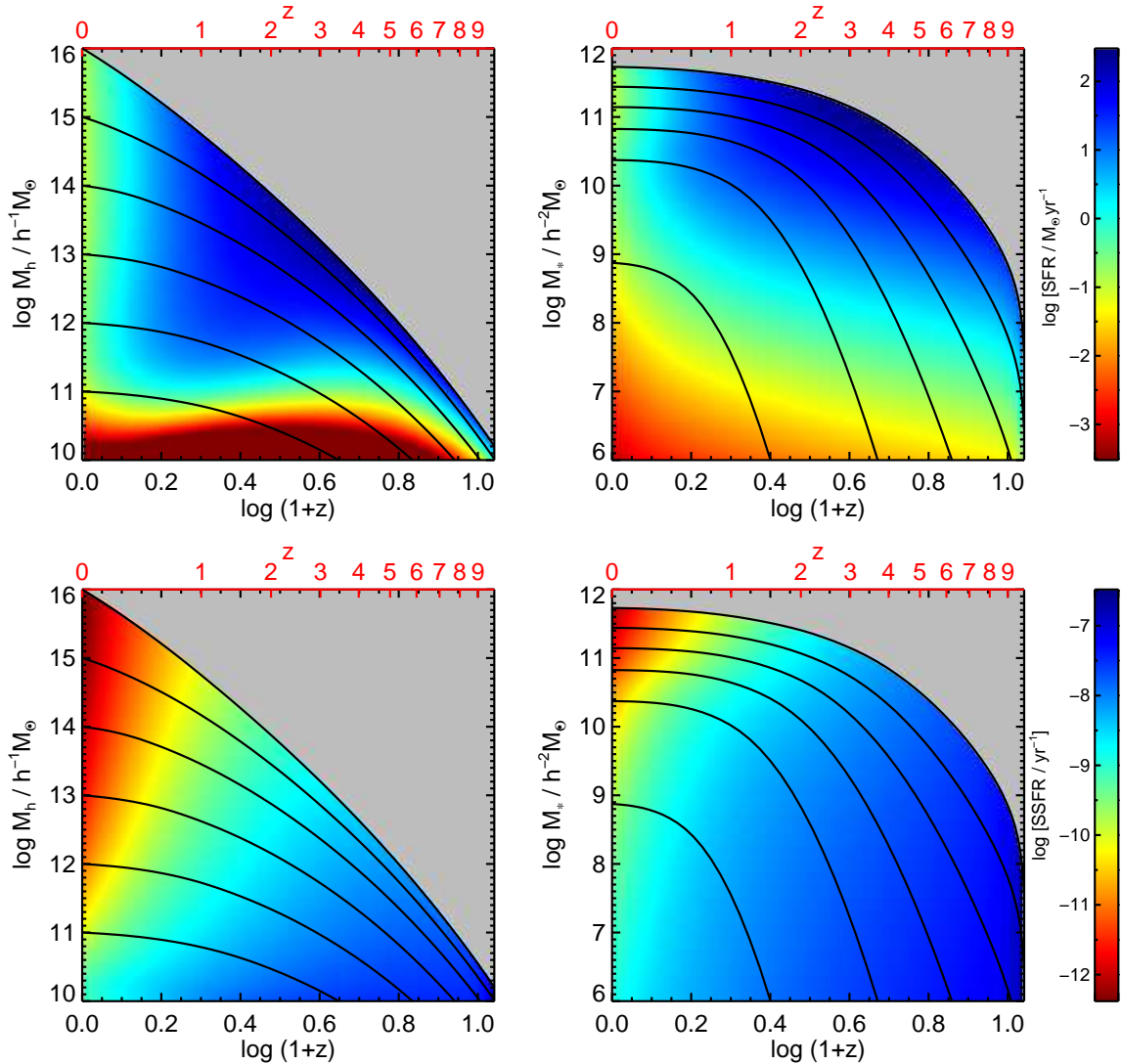}
\caption{Maps  of our  model predictions  for the  star formation  rates (SFR;
  upper  panels) and  specific star  formation rates  (SSFR; lower  panels) of
  central galaxies  as functions  of halo mass  (left-hand panels)  or stellar
  mass (right-hand panels)  and redshift. The vertical bars  on the right-hand
  side indicate  the corresponding  color-coding, where the  SFR and  SSFR are
  expressed  in   units  of  $\msun   {\rm  yr}^{-1}$  and   ${\rm  yr}^{-1}$,
  respectively.  The solid curves in each of the panels show the median growth
  of halo mass  (left-hand panels) and stellar mass  (right-hand panels) along
  the main branch of halos with different present-day masses.}
\label{fig:SFR_map}
\end{figure*}

\subsection{Comparison with some recent results}
\label{sec:comparison}

As pointed out in the introduction, there are a number of recent studies that,
similar to this paper, modelled the SFHs of galaxies as function of halo mass.
All these studies  make use of SHAM (or a  modified version thereof), together
with halo merger trees generated  from N-body simulations, to make predictions
for the  mass growth and  SFH of galaxies  (e.g., Moster \etal  2013; Behroozi
\etal 2012, 2013;  Wang \etal 2012).  In all  these analyses, assumptions have
to  be made  regarding  the  evolution of  satellite  galaxies, in  particular
regarding  the contribution  of satellite  accretion to  the mass  assembly of
central  galaxies.    Although  the  results  from  all   these  analyses  are
qualitatively similar to each other and to our results, there are substantial,
quantitative  differences.   In  this   subsection,  we  make  a  quantitative
comparison between our results and those obtained by Moster \etal (2013).

Fig. \ref{fig:SFH_comp}  shows our model  predictions of the SFHs  for central
galaxies. The  lower and upper  boundaries of the  shaded areas are  our model
predictions  based  on  SMF1 (the  same  as  the  long  dashed lines  in  Fig.
\ref{fig:SFH})   and  SMF2   (the   same   as  the   long   dashed  lines   in
Fig. \ref{fig:SFH_2}), respectively.  The long-dashed line shown in each panel
is the  result obtained  with the  fitting formula provided  by Moster  et al.
(2013).  The Moster et al.  results are obtained by assuming a WMAP7 cosmology
and a Chabrier (2003) IMF; these are sufficiently similar to the cosmology and
IMF  adopted  here,  that  a  direct  comparison  is  justified.   Within  the
uncertainty  ($\sim 0.2\,{\rm  dex}$ in  the SFH),  the peaks  in the  SFHs of
Moster   et   al.    (2013)    are   roughly   consistent   with   our   model
predictions. However,  at high-$z$  their SFHs are  higher than ours  while at
low-$z$ and for massive halos, the Moster et al. SFHs are lower than our model
predictions.   These differences,  especially  those at  low-$z$, most  likely
comes  from  the  different   treatment  of  satellite  galaxies.   Moster  et
al. (2013) assume that 20\% of the stellar mass of accreted satellite galaxies
`escapes' from the central galaxy, ending  up as diffuse halo stars. If we add
such a prescription  to our model, we predict a low-$z$  behavior that is very
similar  to  that  of Moster  et  al.,  although  we  find that  a  `satellite
disruption fraction' of  $\sim 45$\%, rather than $20$\%,  yields results that
are in better agreement with observational constraints on the SFHs.

It is also interesting to compare  our results to those of Leitner \& Kravtsov
(2011) and Leitner (2012), who modeled the stellar mass growth in star-forming
galaxies  using   the  observed   SFR-stellar  mass  relations   at  different
redshifts. The short-dashed and  dotted lines in Fig.\,\ref{fig:SFH_comp} show
the  model predictions  of  Leitner  \& Kravtsov  (2011)  and Leitner  (2012),
respectively.  Although their  models predict SFH peaks that  are very similar
to our  model predictions, their models  predict SFRs at  low (high) redshifts
that are significantly higher (lower).  The difference at low-$z$ is expected,
because their models only describe star-forming galaxies, while our models are
for the  general population. At high-$z$,  the results of  Leitner \& Kravtsov
(2011) and  Leitner (2012) are  not obtained directly from  observational data
but rather  from extrapolations of the  low-$z$ data to  higher redshifts.  We
therefore suspect that the difference  between their results and ours reflects
an amplification of  the uncertainties in their SFR-stellar  mass relations at
lower $z$.

\begin{figure*}
\plotone{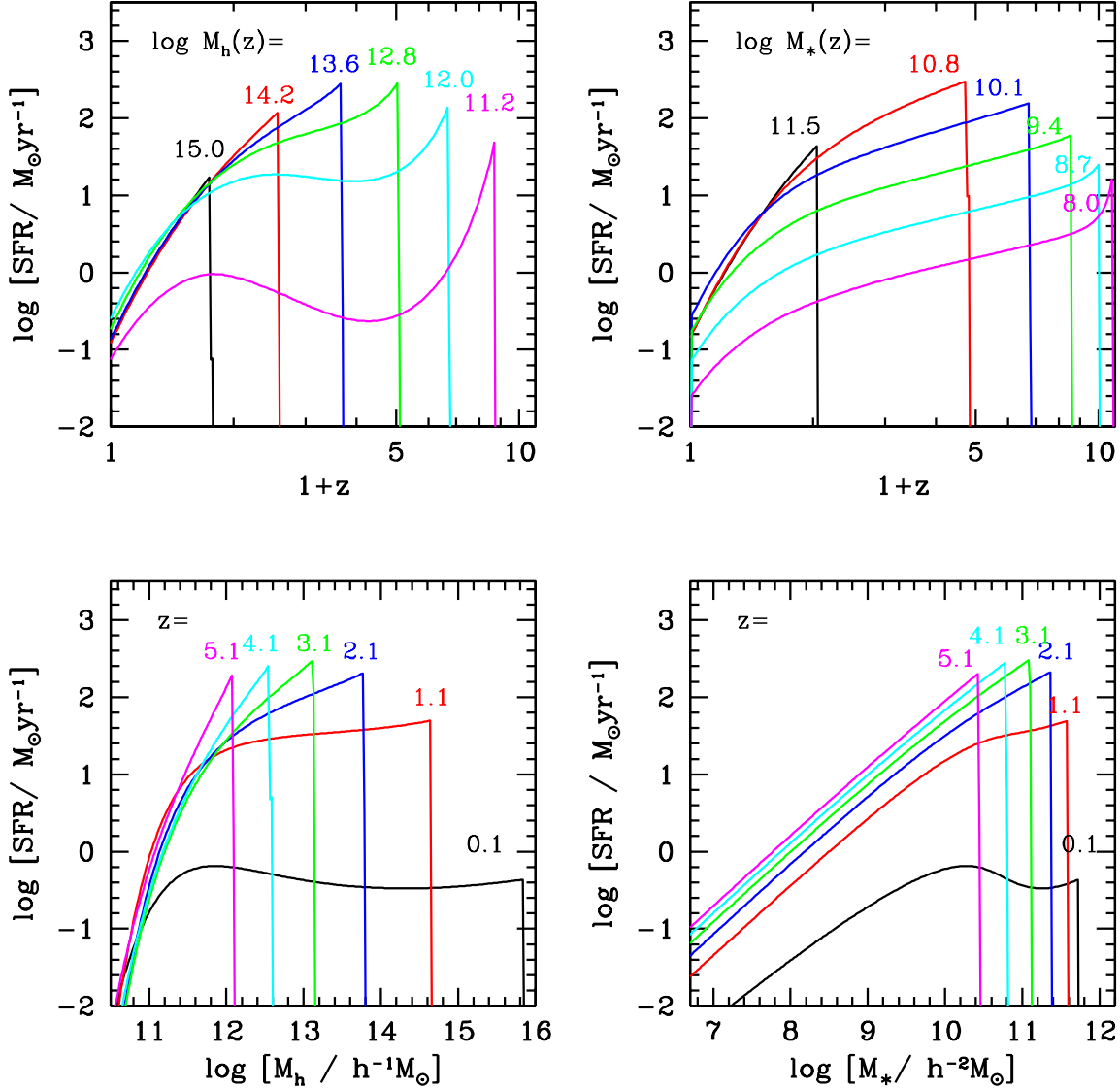}
\caption{Model predictions  for the star formation rates  of central galaxies.
  The upper panels  show the SFRs of central galaxies  as function of redshift
  for  different halo  masses (upper  left-hand panel)  and  different stellar
  masses (upper right-hand panel), as labelled. The lower panels show the SFRs
  of central galaxies as functions  of halo mass (left-hand panel) and stellar
  mass  (right-hand panel)  at  different redshifts,  as  labelled. All  these
  curves are either vertical or horizontal  cuts through the SFR maps shown in
  the upper panels of Fig.~\ref{fig:SFR_map}.  In each of the four panels, the
  vertical cut-offs in  the curves correspond to median main  branch mass of a
  halo with present day mass $M_h = 10^{16}\msunh$.}
\label{fig:SFR_xx}
\end{figure*}
\begin{figure*}
\plotone{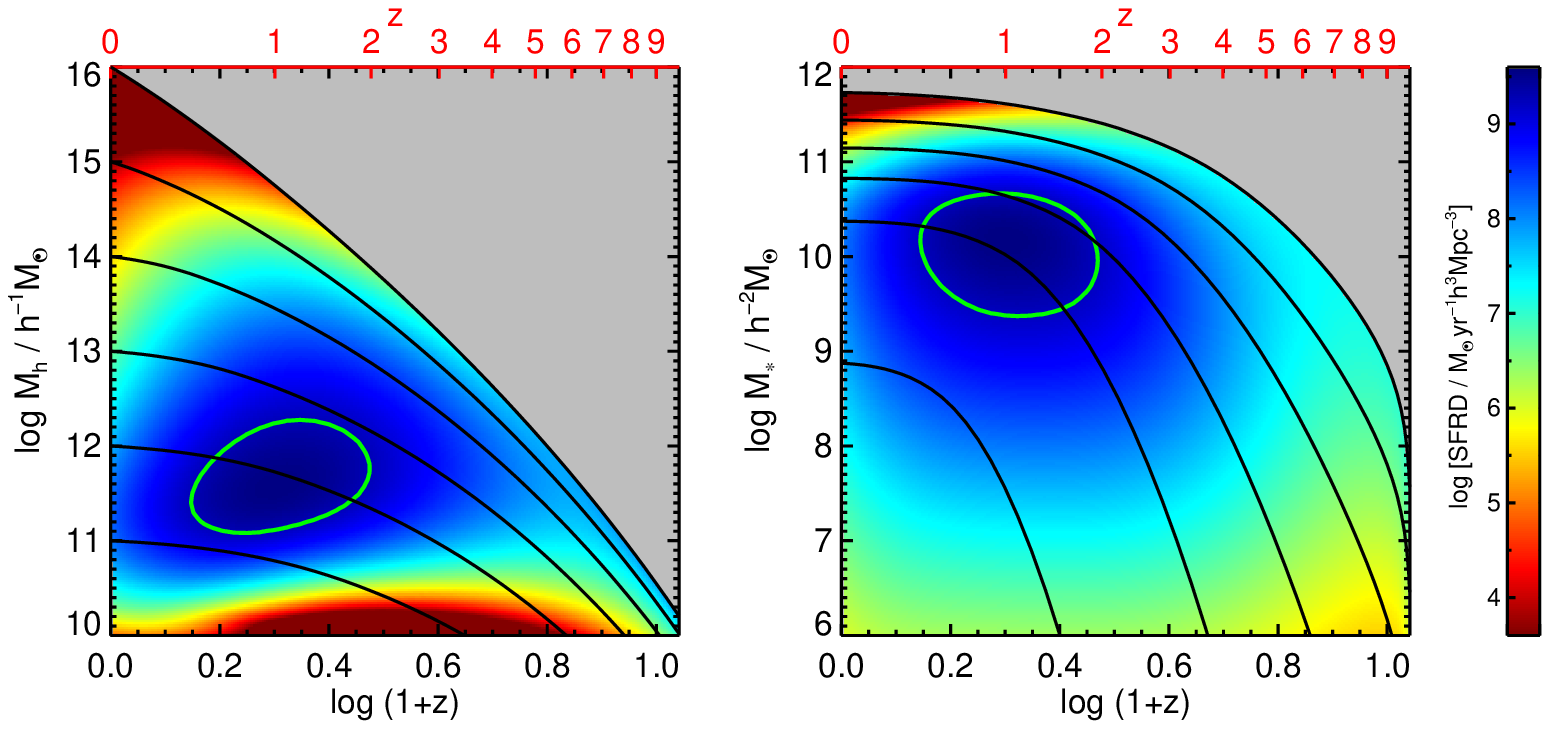}
\caption{Maps of  our model  predictions for the  star formation  rate density
  (SFRD)  of central  galaxies.  The  color coding  used is  indicated  in the
  vertical bar  on the right-hand site.  These SFRDs are normalized  to be the
  mass  density of  stars formed  per unit  $\log M_h$  per unit  $\log (1+z)$
  (left-hand  panel)  or  per  unit   $\log  M_\ast$  per  unit  $\log  (1+z)$
  (right-hand  panel),  and are  expressed  in  units  of $\msun{\rm  yr}^{-1}
  [\mpch]^{-3}$.   The  contours in  the  two  panels  enclose a  region  that
  contributes  50\%  of  the  total  stars  formed in  the  universe.   As  in
  Fig.~\ref{fig:SFR_map}, the solid curves show the median growth of halo mass
  (left-hand panel) and stellar mass  (right-hand panel) along the main branch
  of halos with different present-day masses.}
\label{fig:SMD_map}
\end{figure*}

\section {Model predictions}\label{sec_predict}

In this section we use our model  for the SFHs of central galaxies as function
of halo  mass to make  a number of  predictions that provide  valuable insight
into how  star formation proceeds in  the Universe. Note  that throughout this
section,  unless specified  otherwise,  our model  predictions  are made  from
observational  constrained  SFHs obtained  from  SMF1  (as  SMF1 shows  better
agreement with the SFR measurements at $z=0.1$ than SMF2).

\subsection{The star formation rate as a function of halo mass and 
redshift} \label{sec:SFD}

Let us  first look at  the SFR maps  in the $M_h$ -  $z$ and $M_{\ast}$  - $z$
spaces,  shown in  the upper  row  of panels  of Fig.~\ref{fig:SFR_map}.   For
reference, the  solid lines show the  growth of the host  halo mass (left-hand
panel)  and the growth  of stellar  mass of  the central  galaxies (right-hand
panel). These  relations are  obtained from the  halo mass accretion  model of
Zhao \etal  (2009) and from the median  stellar mass - halo  mass relation for
central     galaxies     obtained     by     Y12     (Eq.\,[\ref{cengrowth}]),
respectively. Results are  only shown for systems that  have $M_h \leq 10^{16}
\msunh$, since more massive halos are extremely rare. As one can see, the SFRs
peak in massive  galaxies with $M_{\ast} \sim 10^{11.0}\msunhh$  in halos with
$M_h  \sim  10^{13.5}\msunh$  at  redshift  $z\sim 2.5$,  and  the  peak  star
formation  rate is about  $100\msun\,{\rm yr}^{-1}$.   At low  redshift ($z\la
0.5$), SFRs are  highest in galaxies with $M_{\ast}  \sim 10^{10.0}\msunhh$ in
halos with $M_h \sim 10^{12}\msunh$.  Also there is a lower halo mass limit at
$\la 10^{10.5}\msunh$, below  which the star formation is  low over the entire
redshift range. Note that these results are in good qualitative agreement with
the results of Behroozi \etal (2012).

In addition to the SFR, in  the lower panels of Fig.~\ref{fig:SFR_map} we show
the corresponding SSFR  maps. In general the SSFR maps  show a smooth gradient
in redshift, being  higher at higher redshift.  At high  $z$, the SSFR depends
only weakly on halo mass and galaxy  mass. Only after the peak redshift of the
SFR (see Eq.   [\ref{equ:SFR_zpeak}]), does the SSFR decrease  more rapidly in
more massive galaxies hosted by more massive halos.

To  better illustrate the  behavior of  the SFRs,  it is  useful to  show some
horizontal  and vertical  cuts  of the  SFR  maps. The  upper  panels of  Fig.
\ref{fig:SFR_xx} show the  SFR as a function of  redshift for central galaxies
in  different  mass halos  (upper-left  panel)  and  for central  galaxies  of
different stellar  masses (upper-right panel).   In the lower panels,  we plot
the  SFR as  a  function of  halo  mass (lower-left  panel)  and stellar  mass
(lower-right panel) at different  redshifts, as indicated.  The sharp cut-offs
in each panel correspond to the median main branch mass of a massive halo with
$M_h =  10^{16}\msunh$ at $z  = 0$, beyond  which the abundance of  systems is
negligibly small.

A number of interesting characteristics  are evident from these plots.  First,
for central galaxies in massive halos  with $M_h \gta 10^{12} \msunh$, the SFR
increases   monotonically  with   redshift  (see   the  upper-left   panel  of
Fig.~\ref{fig:SFR_xx}). In  less massive halos, however,  the SFR-$z$ relation
is not  monotonic. Rather it  reveals two maxima,  one at high redshift  ($z >
5$), and one at intermediate redshift $z \sim 1$. Contrary to the SFRs in more
massive halos,  the SFR-$z$ relation  has a local  minimum at a  redshift that
increases with decreasing  halo mass. However, we causion  that such 
 two maxima  features in small halos are  obtained from the extrapolations
  of the current  observational stellar mass and  redshift limits  (see left
  panel of Fig.  \ref{fig:SFH_peak}).  Second, for central  galaxies of given
stellar mass the SFRs increase monotonically with increasing redshift over the
entire  redshift  ranges probed,  with  a weak  indication  that  the rate  of
increase  is  larger for  more  massive  centrals  (see upper-right  panel  of
Fig.~\ref{fig:SFR_xx}).  Third, at $z\ga 2$ the SFR is higher for more massive
halos  (see the  lower-left  panel of  Fig.~\ref{fig:SFR_xx}).   At $z\la  2$,
though, the SFRs in massive halos are strongly suppressed relative to those at
higher $z$. The  mass scale at which this  suppression becomes apparent shifts
to lower halo  masses with decreasing redshift; for small  halos with $M_h \la
10^{11.2} \msunh$, the SFR is  almost independent of redshift.  Finally, as is
apparent from  the lower right-hand  panel of Fig.~\ref{fig:SFR_xx},  the SFRs
for galaxies  of a given stellar  mass generally increase  with redshift, and,
for a given redshift, the SFR  roughly shows a power-law dependence on stellar
mass, especially  at high  redshift. These features  are quite robust  for the
observationally constrained SFHs, regardless whether they are constrained from
SMF1 or SMF2 (see Appendix A).

\begin{figure*}
\plotone{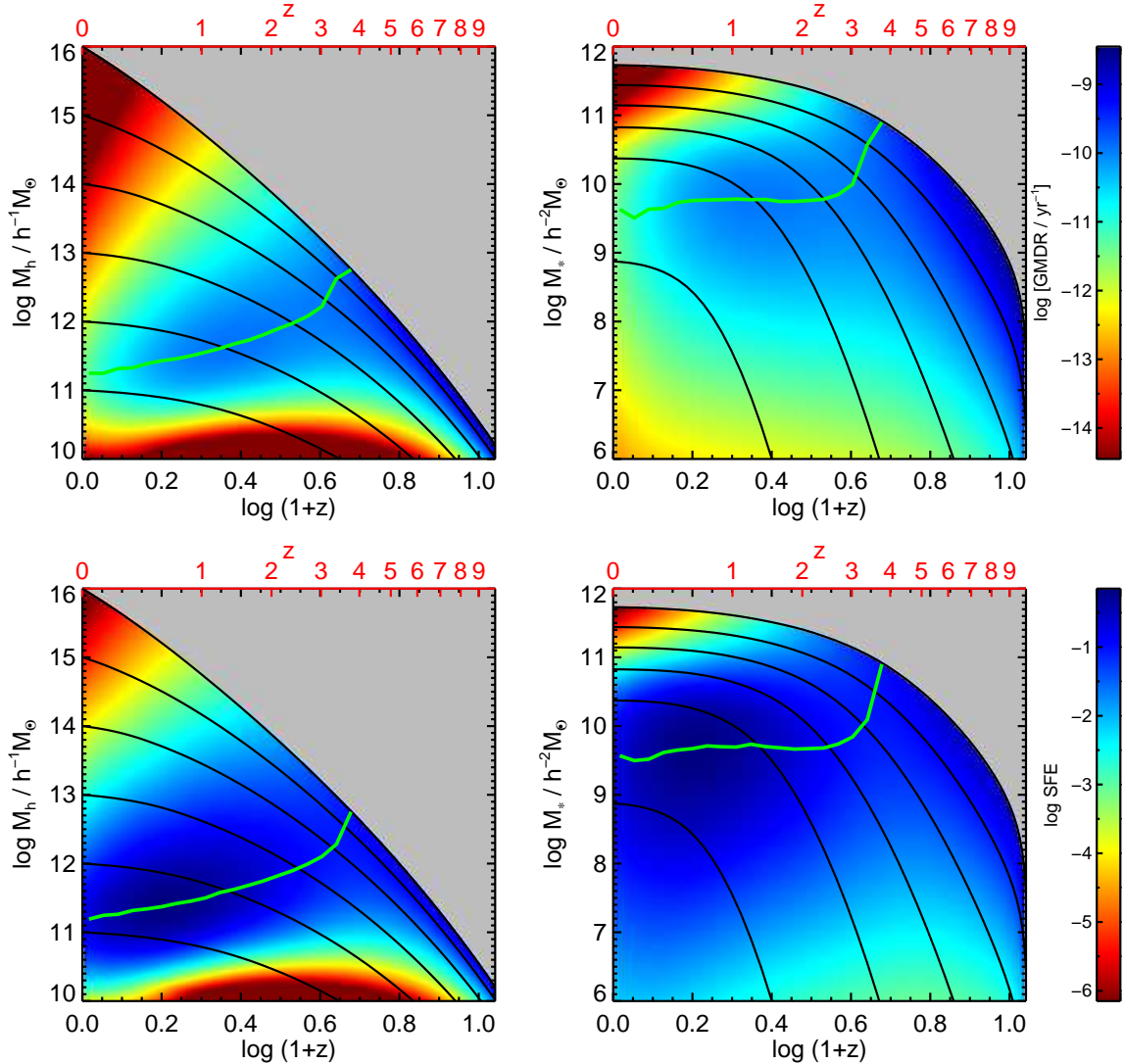}
\caption{Model  predictions for  the  gas mass  depletion  rates (GMDR;  upper
  panels)  and star  formation  efficiencies (SFE;  lower  panels) of  central
  galaxies  as functions  of  halo  mass (left-hand  panels)  or stellar  mass
  (right-hand panels) and  redshift.  The color codings used  are indicated in
  the  vertical   bars  on  the   right-hand  site.   Whereas  the   SFEs  are
  dimensionless,  the  GMDRs   are  in  units  of  ${\rm   yr}^{-1}$.   As  in
  Fig.~\ref{fig:SFR_map}, the  solid, black curves  show the median  growth of
  halo mass  (left-hand panel) and  stellar mass (right-hand panel)  along the
  main branch  of halos  with different present-day  masses, while  the solid,
  green lines mark the maximum GMDRs/SFEs as function of redshift. }
\label{fig:SFE_map}
\end{figure*}

\subsection{Star formation rate density}
\label{sec:SMD}

Using the `SFR  maps' shown in the upper  panels of Fig.~\ref{fig:SFR_map}, we
can obtain similar maps but for  the star formation rate density (SFRD). These
have the advantage that they highlight when and where the majority of stars in
the Universe formed. The SFRD is related to the SFRs according to 
\begin{eqnarray}\label{equ:SMD_map1}
{\rm SFRD}(M_h,z) &=&  \,{\rm ln} 10\times n(M_h, z) \,  M_h  \\
      &\times& {\rm SFR}(M_h, z) 
{\frac {{\rm d} t}{{\rm d} \log (1+z)}}\,. \nonumber
\end{eqnarray}
where $n(M_h,z) \equiv {\rm d}n(z)/{\rm  d}M_h$ is the comoving number density
of  halos at  $z$ with  masses in  the range  $[M_h,M_h +  {\rm  d}M_h]$. Thus
defined,  the SFRD is  the mass  density of  stars formed  per $\log  M_h$ per
$\log(1+z)$,   and  the   result  is   shown   in  the   left-hand  panel   of
Fig.~\ref{fig:SMD_map}.  The SFRD peaks at redshift $z\sim 1$ in halos of $M_h
\sim 10^{12} \msunh$.   The solid, green contour shown on top  of the SFRD map
encloses the  region that contributes 50\%  to the total SFH  in the Universe,
and shows that half  of the total stellar mass is formed  in halos with masses
in  the range  $10^{11.1}{\msunh} \la  M_h \la  10^{12.3}{\msunh}$ and  in the
redshift  range $0.4  \la  z \la  1.9$.   Overall, halos  with  masses $M_h  =
10^{10.5-13.5} \msunh$ contribute  the vast majority of all  star formation in
the universe.  At $z\la 1.5$, the star formation density becomes progressively
more dominated by low-mass halos.

A SFRD map can also be constructed in the stellar mass - redshift plane,
\begin{equation}\label{equ:SMD_map2}
{\rm SFRD}(M_{\ast},z) = {\rm SFRD}(M_h,z) 
{\frac {{\rm d} \log M_h}{{\rm d} \log M_{\ast}}}\,,
\end{equation}
and the results are shown in the right-hand panel of
Fig.~\ref{fig:SMD_map}.  Here we see that the SFRD peaks at $M_{\ast}
\sim 10^{10} \msunhh$, and that half of the stars in the Universe is
formed in galaxies within a narrow ($\sim 1.3 {\rm dex}$) stellar mass
range: $10^{9.4}{\msunhh} \la M_{\ast} \la 10^{10.7} {\msunhh}$.

\subsection{Star formation efficiency and gas mass depletion rate}
\label{sec:SFE}

Next we  look at the  gas mass depletion  rates (GMDR) of central  galaxies in
halos of  different masses.  Since our  current modeling does  not include any
gas  components, we define  the GMDR  to be  the SFR  normalized by  the total
baryonic mass,  defined as the halo  mass times the  universal baryon fraction
$f_b$: 
\begin{equation}\label{equ:SFE_map1}
{\rm GMDR}(M_h,z) = {{\rm SFR}(M_h,z) \over f_b \, M_h}\,,
\end{equation} 
where  we adopt  $f_b \equiv  \Omega_\rmb /  \Omega_\rmm \simeq  0.167$ (e.g.,
Komatsu \etal 2011).  Thus defined, the  GMDR is the reciprocal of the time it
takes a galaxy to consume all the  gas associated with its halo at the current
SFR.  The  top left and right  panels of Fig.~\ref{fig:SFE_map}  show the GMDR
maps in the halo mass {\it  vs.}  redshift and stellar mass {\it vs.} redshift
spaces,  respectively. The  green solid  line shown  in each  panel  marks the
maximum GMDRs as function of  redshift.  Three trends are worth noting. First,
the GMDRs  in low mass halos  ($M_h \lta 10^{11} \msunh$)  are always strongly
suppressed with respect to their more massive counterparts.  Second, at $z \ga
3.5$,  the  most massive  halos  have the  highest  GMDRs,  which are  roughly
constant  at  $\sim   10^{-9.2}  \,{\rm  yr}^{-1}$  over  a   large  range  in
redshift. These high GMDRs result in  rapid growth of the central galaxies, as
is evident from the black solid  lines shown in the right-hand panel. Finally,
below $z \sim 3.5$ `downsizing' kicks in,  in that the peak in the GMDR shifts
to  lower halo  masses with  decreasing redshift.   There appears  to  be some
`quenching' mechanism  (or at  least some mechanism  that manages  to strongly
suppress  the GMDR),  which  operates in  halos  whose mass  shifts down  with
decreasing  redshift; centrals  that end  up in  more massive  halos  at $z=0$
quench at a higher redshift, when their halo mass is more massive. Whereas the
present-day quenching  mass is close  to $10^{11} \msunh$, centrals  that have
ended up  in the most  massive halos quenched  around $z \sim 3.5$  when their
main progenitor halo had a mass $\sim 10^{13} \msunh$.

\begin{figure}
\plotone{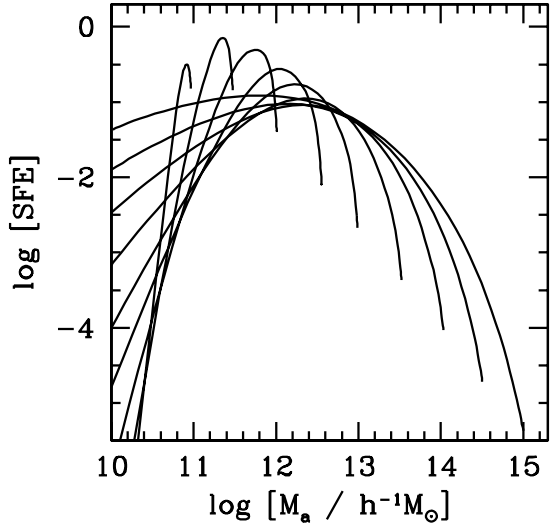}
\caption{Model  predictions for  the  star formation  efficiencies of  central
  galaxies as  function of the  main branch mass,  $M_a$, of their  host halo.
  Results are shown for host  halos with present-day masses $\log (M_h/\msunh)
  =11.0, 11.5, 12.0$,  $12.5, 13.0, 13.5, 14.0, 14.5,  15.0$.  The present-day
  mass for each curve is evident from its end-point at the high-mass end.}
\label{fig:SFE_quench}
\end{figure}

An insightful  way of expressing  the SFHs of  central galaxies is  to compare
them with the mass accretion histories  of their halos. To this end, we follow
Behroozi  et al.  (2013) and  define the  star formation  efficiency  (SFE) of
central galaxies in the main branch progenitors as 
\begin{equation}\label{equ:SSFE}
{\rm SFE}(M_h,z) = \frac{{\rm SFR}(M_h,z)}{\rmd (f_b M_a)/\rmd t} \,.
\end{equation} 
Thus  defined the  SFE describes  the  SFR in  the main  branch progenitor  at
redshift $z$  (where the  halo mass is  $M_a$) in  units of the  baryonic mass
accretion  rate at  the  same  redshift\footnote{Note that  we  are unable  to
  distinguish whether  star formation at a given  redshift consumes previously
  or newly accreted  gas.}.  The lower panels of  Fig.  \ref{fig:SFE_map} show
maps of the SFE in the  halo mass {\it vs.}  redshift (left-panel) and stellar
mass {\it  vs.}  redshift  (right-panel) spaces. The  SFE peaks in  halos with
mass $\sim 10^{11.4}\msunh$  at redshift $\sim 0.5$, with  central galaxy mass
$\sim 10^{9.5}\msunhh$. Compared with the GMDR maps shown in the upper panels,
the peak in the SFE shifts to lower halo/stellar mass and lower redshift. This
is simply  a reflection of  the fact that  low mass halos have  lower specific
mass accretion rates than their  more massive counterparts (i.e., more massive
halos assemble later).

These results are in qualitative  agreement with those obtained by Behroozi et
al.   (2013),  but  quantitatively  there  are  differences.   In  particular,
compared to Behroozi et al.   (2013), our model seems to predict significantly
stronger redshift  dependence. For  example, according to  our model  the halo
mass at which the SFE is  highest increases from $M_h\sim 10^{11.2} \msunh$ at
$z =  0$ to $M_h \sim  10^{12.1}\msunh$ at $z =  3$ (see solid,  green line in
lower  left-hand   panel  of  Fig.   \ref{fig:SFE_map}).    In  contrast,  the
corresponding ${\rm SFE}$ peak halo masses obtained by Behroozi et al.  (2013)
are $10^{11.5} \msunh$  and $10^{12.0}$ at $z = 0$ and  $z = 3$, respectively.
Interestingly, we find that the {\it stellar mass} of the central galaxies for
which the  SFE (or GMDR)  is maximal is  virtually independent of  redshift at
$M_{\ast,  c}\sim  10^{9.7}\msunhh$,  but  only  for $z\la  2.5$.   Our  model
suggests  that  for  $z\ga  2.5$,  this characteristic  stellar  mass  rapidly
increases to  effectively become  equal to that  of the most  massive galaxies
present at those redshifts.  Such  a feature is reconfirmed very recently
  using a different approach by Lu et al. (2013 in preparation).

To highlight some of the features of the SFEs, Fig.~\ref{fig:SFE_quench} shows
cuts of the SFE  maps along some of the main branch  histories (solid lines in
the  left panels  of Fig.~\ref{fig:SFE_map}).  Results are  shown  for central
galaxies with present-day halo masses of $\log (M_h/\msunh) =11.0, 11.5, 12.0,
12.5, 13.0,  13.5, 14.0,  14.5, 15.0$. As  the main progenitor  mass increases
with  time, the SFE  initially increases,  reaches a  maximum, after  which it
decreases rapidly to a quenched state.  Interestingly, the initial increase of
the SFE with increasing $M_a$ is much steeper for central galaxies that end up
in  less massive halos.  A rapid  increase of  SFE over  the range  $10^{10} -
10^{11} \msunh$  in $M_a$, which is  predicted by our model  for centrals that
end-up  in present-day halos  with $M_h  \lta 10^{12.5}\msunh$,  is consistent
with the  presence of a `halo  mass floor' $M_{\rm min}  \sim 10^{11} \msunh$,
below which star formation is strongly suppressed, as suggested by Bouch\'e et
al (2010).  However, our model predicts  that central galaxies that  end up in
more massive  halos have fairly high  SFEs when their main  progenitor mass is
$\sim 10^{10} \msunh$. This suggests that  this halo mass floor must have been
substantially lower (or absent) at higher redshifts ($z \ga 5$).

Another    interesting     feature    of    the     SFE($M_a$)    curves    in
Fig.~\ref{fig:SFE_quench} is  the evolution in the `quenching  mass', which we
define as  the main-progenitor mass at  which the SFE  is maximal. Present-day
halos with $M_h\ga 10^{13} \msunh$ all seem to quench when $M_a \sim 10^{12.5}
\msunh$, at which point their SFE is $\sim 0.1$ (i.e., the star formation rate
is ten  percent of  the baryon  accretion rate) .  For present-day  halos with
$M_h\la 10^{13}  \msunh$ our model  predicts a `downsizing' behavior,  in that
the quenching mass  shifts to lower $M_a$ for present-day  halos that are less
massive.  In  addition, the peak value  of the SFE increases,  coming close to
unity for $M_h \sim 10^{11.5} \msunh$.

An  important,  outstanding question  in  galaxy  formation  is what  physical
process is responsible  for the quenching of central  galaxies, which seems to
happen whenever  the halo  mass is of  order $10^{12}  \msunh$. Interestingly,
this mass scale is very similar  to the one that separates cold-mode accretion
and  hot-mode  accretion  (Birnboim  \&  Dekel 2003;  Kere\v{s}  \etal  2005),
suggesting that  the quenching  of star formation  in central galaxies  may be
related to the ability of a dark matter halo to form a hot gaseous halo.  This
is indeed  what seems to be needed  to explain the observed  bimodality in the
distributions  of  galaxy colors  and  specific  star  formation rates  (e.g.,
Cattaneo \etal 2006; Birnboim \etal  2007).  In addition to cold mode/hot mode
accretion, other mechanisms have also been invoked to explain the quenching of
star formation in  massive central galaxies, ranging from  AGN feedback (e.g.,
Tabor \& Binney 1993; Ciotti \&  Ostriker 1997; Croton \etal 2006; Bower \etal
2006;  Hopkins  \etal 2006)  and  gravitational  heating  (e.g., Fabian  2003;
Khochfar \& Ostriker 2008; Dekel \&  Birnboim 2008; Birnboim \& Dekel 2011) to
thermal conduction (e.g.,  Kim \& Narayan 2003).  Although  it remains unclear
which of these processes dominates, and  how exactly they operate, it is clear
that any  successful model  has to be  able to  explain why star  formation in
galaxies is quenched once their halo masses reach a characteristic mass, which
`downsizes' from $\sim  10^{12.5} \msunh$ at high redshifts  ($z \gta 3.5$) to
$\sim 10^{11.5}$ at the present-day.

\subsection{Comparing SFE with central-to-host halo mass ratios} 
\label{sec:ratio}

\begin{figure*}
\plotone{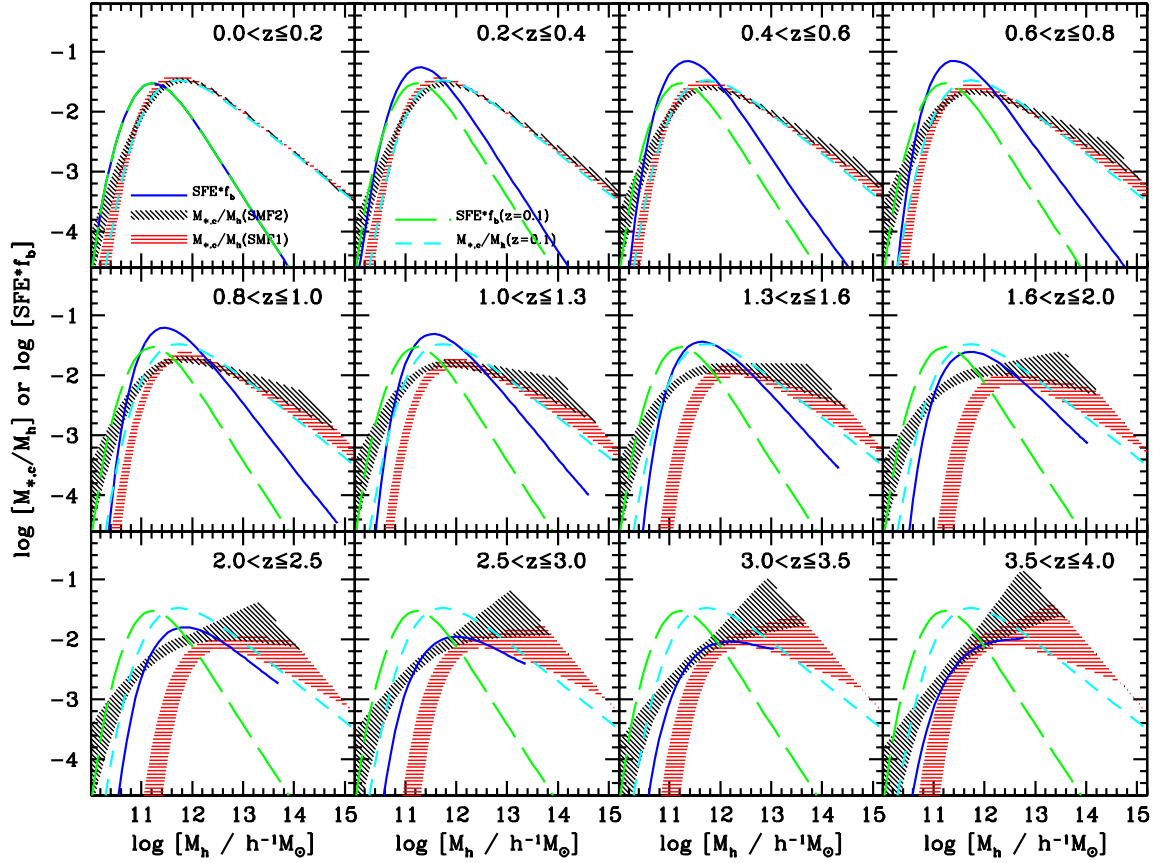}
\caption{Instantaneous versus integrated mass ratios. The red and black shaded
  regions indicate the 68\% confidence  levels on the `integrated' mass ratios
  between central galaxy and host  halo, $M_{\ast,c}/M_h$, as function of host
  halo  mass,  as  obtained by  Y12  from  the  SMF1  and SMF2  data  samples,
  respectively. Different  panels correspond to  different redshift intervals,
  as  indicated. The solid  curve in  each panel  is the  `instantaneous' mass
  ratio, ${\rm SFE}\times  f_b$, inferred from our model.   For comparison, we
  also  show in each  panel the  model predictions  assuming ``no-evolution'',
  which simply are the  $M_{\ast,c}/M_h$ (cyan, short-dashed curves) and ${\rm
    SFE}\times f_b$ (green, long-dashed curves) at $z=0.1$.}
\label{fig:SFE_ratio}
\end{figure*}

Since the SFE indicates the fraction of newly accreted baryonic matter that is
converted  into stars, the  quantity SFE$\times  f_b$ can  be regarded  as the
instantaneous mass  ratio between  the newly formed  stars and  newly accreted
dark matter. It will be interesting to compare this mass ratio to that between
the stellar  mass of a  central galaxy  and the mass  of its dark  matter host
halo.  The latter  has been  extensively studied  in recent  years, and  is an
integration of SFE$\times f_b$ over cosmic time plus a contribution due to the
accretion of satellite  galaxies (important in massive halos  at low redshift)
and the  impact of passive evolution  (which removes $\sim 40$  percent of the
stars formed at early times from the mass budget of present day stars).

We  first  show,  in  Fig.~\ref{fig:SFE_ratio},  the  $M_{\ast,c}/M_h$  ratios
obtained  in  Y12  for  SMF1  (red  shaded  curves)  and  SMF2  (black  shaded
curves). The  shaded areas reflect  the 68\% confidence levels  resulting from
the  statistical  errors  in   the  resulting  $M_{\ast,c}/M_h$  ratios.   The
differences  between the  SMF1  and SMF2  curves  can be  regarded as  roughly
reflecting  the  systematical errors  in  the  current  data. Compared  to  the
"no-evolution" model,  which simply is the $M_{\ast,c}/M_h$  at $z=0.1$ (cyan,
dashed  curve   in  each  panel),  the  $M_{\ast,c}/M_h$   ratio  does  evolve
significantly, especially beyond  redshift $z \simeq 1$. On  average, the peak
$M_{\ast,c}/M_h$ ratio at  redshift $z\ga 2.0$ is $\sim  0.5$dex below the one
at redshift $z = 0.1$.

The solid line  in each panel of Fig.~\ref{fig:SFE_ratio}  indicates our model
prediction for  the {\it instantaneous}  ratio, ${\rm SFE}\times f_b$,  at the
corresponding redshifts,  reduced by  40\% in order  to (roughly)  correct for
passive  evolution   (see  discussion  in   \S\ref{sec:growth}).   Again,  for
comparison, we also  show in each panel the  ``no-evolution" model predictions
(long-dashed green curve), which simply  is the ${\rm SFE}\times f_b$ ratio at
$z=0.1$. Note how the peak of the ${\rm SFE}\times f_b$ {\it vs.}  $M_h$ curve
first increases by about $+0.4$dex from $z=0.1$ to $z \simeq 0.6$, after which
it gradually  decreases to about  $-0.5$dex at $z\ga  3.0$.

Comparing   the   `integrated'    mass   ratios,   $M_{\ast,c}/M_h$   to   the
`instantaneous' ratios,  ${\rm SFE} \times  f_b$, in different  redshift bins,
one notices that the two ratios are  in good agreement with each other at high
redshifts ($z \ga  3$). This is as  expected, since (i) the SFE  does not show
strong  time  evolution at  high  redshift,  and  (ii) the  contribution  from
satellite galaxies  is negligible. Moving to  lower redshifts, we  see an ever
increasing `lag'  between the `integrated' and `instantaneous'  mass ratios at
the  high mass  end.  This  is  a manifestation  of the  quenching of  central
galaxies in massive  haloes; their instantaneous SFRs are  a poor indicator of
their integrated (past) SFHs.  In addition, massive centrals may have accreted
a significant fraction of their stellar mass in the form of satellite galaxies
(see  \S\ref{sec:situ}   below),  which  is   also  not  reflected   in  their
instantaneous  SFRs.  Finally,  at all  $0.2\la z  \la 2.5$,  the peak  in the
`instantaneous'  mass ratio  is  higher  than that  in  the `integrated'  mass
ratio. This  is a manifestation of the  fact that the instantaneous  SFE has a
global  peak at  redshift $z\sim  0.5$.  Hence,  at redshifts  above and  in a
certain range blow this peak  redshift, the instantaneous mass ratio should be
larger than its time-integrated equivalent.

\subsection{The cosmic star formation densities}
\label{sec:CSFD}

\begin{figure*}
\plotone{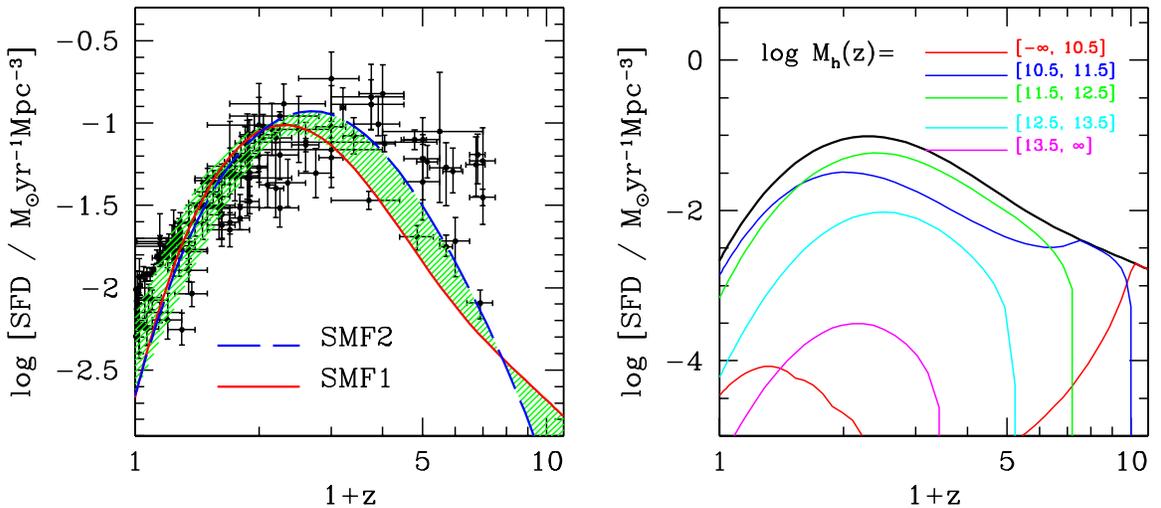}
\caption{The  cosmic   star  formation  densities.    {\it  Left-hand  panel:}
  comparison of our model predictions  with the observational data compiled by
  Hopkins \& Beacom (2006; small dots with error bars).  Our model predictions
  based  on SMF1  and  SMF2 are  shown  as the  solid  and long-dashed  lines,
  respectively.  For comparison,  the shaded band indicates the  range of SFDs
  obtained from the `MIN', `MAX' and  `OBS' models based on SMF1 and SMF2, and
  therefore  (roughly)  reflects  our  model uncertainties.   {\it  Right-hand
    panel:} model predictions based on SMF1 for the contributions to the total
  SFD due to halos of different masses, as indicated.}
\label{fig:CSFD}
\end{figure*}

Once we know how the SFRs of  central galaxies depend on $M_h$ and $z$, we can
combine  the  dependence with  the  halo mass  function  to  predict the  star
formation history of  the universe, as described by  the cosmic star formation
density (SFD) defined as 
\begin{equation}\label{equ:CSFD}
{\rm SFD}(z) = \int^{\infty}_{0}{\rm SFR}(M_h, z)~n(M_h, z) {\rm d} M_h \,.
\end{equation}
Note  that our  model prediction  only accounts  for the  contribution  due to
central  galaxies, whereas  the  data  on the  cosmic  star formation  history
includes  contributions  from both  centrals  and  satellites. However,  since
satellite  galaxies contribute  less  than  $\sim 40\%$  of  the total  galaxy
population  (e.g., Mandelbaum  \etal 2006;  Tinker \etal  2007; van  den Bosch
\etal 2007; Yang \etal 2008; Cacciato \etal 2013) and have significantly lower
SFRs (e.g.,  Balogh \etal 2000; Weinmann  \etal 2006, 2009;  Wetzel, Tinker \&
Conroy 2012a,b),  their contribution to the  SFD is sufficiently  small that a
comparison of our model prediction with data is still meaningful.

In the left  panel of Fig.~\ref{fig:CSFD}, we compare  our model prediction of
the cosmic SFD  to a compilation of data taken from  Hopkins \& Beacom (2006).
Hopkins \&  Beacom (2006)  converted all  the SFRs to  a Salpeter  (1955) IMF,
which is different from the Kroupa (2001) IMF used in this work.  As suggested
in  Perez-Gonzalez \etal  (2008), the  stellar  masses based  on the  Salpeter
(1955) IMF  are systematically  larger by  a factor of  $\sim 1.7$  than those
based  on the  Kroupa  IMF. Here  we have  applied  such a  conversion in  our
comparison by multiplying  our model prediction by a  factor $1.7$.  The solid
and  long dashed  lines show  our model  predictions based  on SMF1  and SMF2,
respectively, with  the latter obtained  from the fitting formula  provided in
Appendix A.   For comparison we also show,  using a shaded band,  the range of
the SFDs obtained  from the three models for the SFR,  `OBS', `MAX' and `MIN',
and  using   SMF1  and  SMF2.   This  band  therefore  roughly   captures  the
uncertainties in our model.

As one can see, at low redshift  ($z\la 2.0$), our model prediction is in good
agreement with the data.  At $z\ga 2.0$, however, our model under-predicts the
cosmic SFD  compared to  the data.  Several  sources might contribute  to this
discrepancy.  First,  our model  prediction of  the SFD is  based on  the {\it
  median} SFH of central galaxies.  In  reality, for a given halo mass the SFR
distribution   has   a    broad   (roughly   log-normal)   distribution   (see
Fig.~\ref{fig:SFR_z0}).  For  a log-normal distribution the  {\it average} SFR
is larger than its {\it median} value by a factor of $e^{(\ln(10)\sigma)^2/2}$
which  corresponds to  an enhancement  in the  SFDs of  $\sim 0.1$dex  for the
typical dispersion, $\sigma\sim  0.3$, shown in Fig.~\ref{fig:SFR_z0}.  Taking
this into account will increase the predicted SFD, especially at $z=0$ because
of   the   bi-lognormal  distribution   of   galaxies   in   halos  of   $\sim
10^{12}\msunh$. Such an  increase of $\sim 0.1$dex is  insufficient to explain
the apparent  discrepancy at  high $z$.  An  alternative explanation  might be
that we  did not include the  contribution due to  satellite galaxies.  Albeit
small,  adding this  contribution will  also slightly  increase our  SFD model
predictions.    In addition to these,  the   high  redshift  SFD  depends
  significantly  on the correction  of faint  galaxies.  As pointed out
  recently  by Behroozi \etal  (2012) based  on some  new measurements  of the
  cosmic  SFD (e.g.,  Bouwens \etal  2012), the  data compiled  by  Hopkins \&
  Beacom  (2006) is likely  to overestimate  the cosmic  SFDs in  the redshift
  range $3  \la z  \la 8$. This  discrepancy is  largely due to  the different
  luminisity cuts  in calculating the SFD  (see Kistler et  al. 2009).  Taking
  all  these uncertainties/issues  into account,  we conclude  that  our model
  seems to  predict a  cosmic SFD somewhat  lower than  the data at  very high
  redshift.  The  discrepancy can  be significantly eased  either if the
  stellar mass  functions at  high redshift have  a significantly  steeper low
  mass end  slope, or  if the faint end slopes of the galaxy luminosity
  functions used in the SFD  measurements are significantly under-estimated.

Finally,  the   right-hand  panel  of  Fig.~\ref{fig:CSFD}   shows  our  model
prediction for  how halos  of different masses  contribute to the  cosmic SFD.
Halos  with  $M_h  <  10^{10.5}  \msunh$  are  predicted  to  only  contribute
significantly at  very high redshifts ($z  \ga 10$). Halos  with masses $10.5<
\log (M_h/\msunh)  <11.5$, on  the other  hand, are predicted  to be  the main
contributors of  the cosmic SFD at  both $z\sim 0$ and  $z\sim 7.0$. Milky-Way
sized halos  with masses $11.5< \log  (M_h/\msunh) <12.5$ are  predicted to be
the main  contributors for most  of the history  of the Universe, all  the way
from $z\sim 7$ to $z \sim  0.1$.  Interestingly, our model predicts that halos
with $M_h  \ga 10^{12.5} \msunh$  never contributed significantly  (i.e., more
than 10\%) to the cosmic star formation density at any redshift.

\subsection{The fraction of stars formed {\it in situ}}
\label{sec:situ}

\begin{figure*}
\plotone{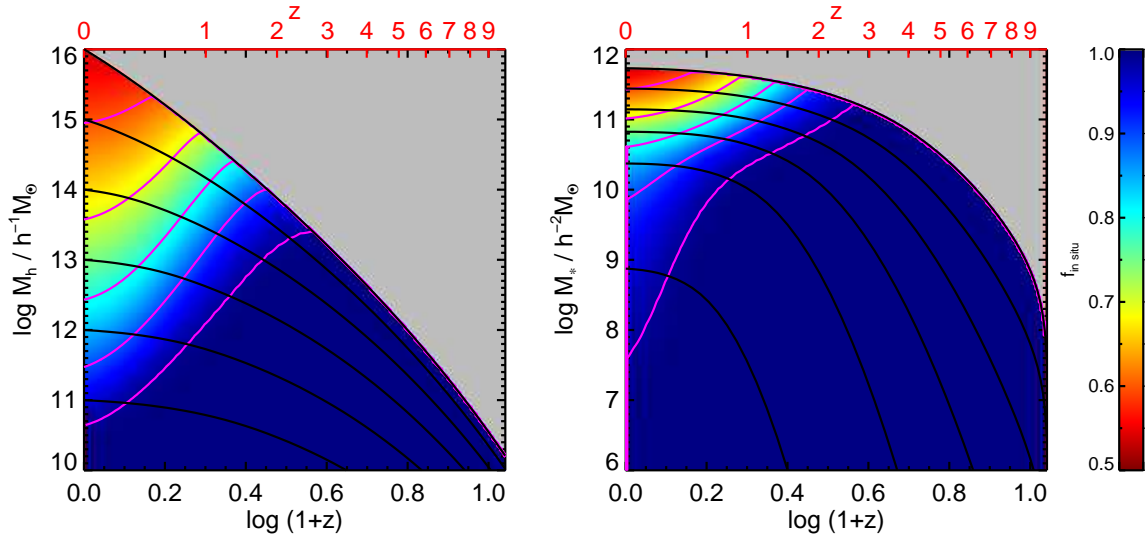}
\caption{Model predictions for the fraction  of stars of central galaxies that
  formed {\it in situ}, $f_{\rm in situ}$, as function of halo mass (left-hand
  panel)  or stellar  mass (right-hand  panel) and  redshift. Color  coding is
  indicated in  vertical bar at the  right-hand side. For  clarity, the purple
  lines indicate the  contours where $f_{\rm in situ}=0.6,  0.7, 0.8, 0.9$ and
  $0.99$, respectively.   The solid  black curves once  again show  the median
  growth of  halo mass (left-hand  panel) and stellar mass  (right-hand panel)
  along the main branch of halos with different present-day masses.}
\label{fig:insitu}
\end{figure*}

Recent cosmological simulations of galaxy  formation show that the assembly of
massive  galaxies,  which are  almost  always  ellipticals,  consists of ``two
phases"; a rapid early phase at $z  \ga 2$, during which stars are formed {\it
  in situ} (i.e.,  within the galaxy) from infalling cold  gas, followed by an
extended phase  during which  {\it ex  situ} stars (in  the form  of satellite
galaxies) are  accreted (e.g. Oser  \etal 2010, 2012; Hirschmann  \etal 2012).
Such a two-phase  formation scenario for massive galaxies  is supported by the
strong evolution in the observed size-mass relation of massive galaxies (e.g.,
Bezanson et  al. 2009,  and references  therein).  Our model  for the  SFHs of
central galaxies can  predict the fraction of stars formed {\it  in situ} as a
function of redshift, halo mass, and stellar mass.

The total stellar  mass of a central  galaxy at a given redshift  $z_0$ can be
obtained from  Eq.~(\ref{equ:M_*01}).  The stellar  mass of stars  that formed
{\it in situ}  can be obtained by replacing ${\rm SFR}_{\rm  MAX}$ by the real
star formation rate ${\rm SFR}_{\rm OBS}$, so that we can write 
\begin{equation}\label{equ:M_*SF}
M_{\ast,{\rm SF}}(z_0) = \int_0^{t_{z_0}} {\rm SFR}_{\rm OBS}(z(t)) \, 
f_{\rm passive}(t_{z_0}-t) \, {\rm d}t \,.
\end{equation}
Fig.~\ref{fig:insitu}  shows  the  ratio  $f_{\rm  in situ}  =  M_{\ast,  {\rm
    SF}}(z_0) / M_{\ast}(z_0)$, between the mass of stars formed {\it in situ}
and the total stellar mass, in both  the $M_h$ - $z$ (left-hand panel) and the
$M_{\ast}$  - $z$  (right-hand  panel) planes.   We  first focus  on the  most
massive galaxies, which are  typically ellipticals in massive halos. According
to our model, at $z > 2.5$ more than $99\%$ of the stars in these galaxies are
formed  {\it in situ}.   This fraction  decreases as  a function  of redshift,
dropping to $\sim 60\%$ at $z=0$.  Thus even today's most massive ellipticals,
for which the  accretion of stars from (satellite) galaxies  is expected to be
most  important, are  predicted to  have the  majority of  their  stellar mass
contributed by early, {\it in  situ} star formation (on average).  For central
galaxies in less massive halos, the  fraction of stars formed {\it in situ} is
even lower.   For a  Milky Way  sized halo of  $M_h\sim 10^{12}\msunh$  at the
present day, more than $80\%$ of the stars are expected to have formed {\it in
  situ}.   Clearly,  major mergers  of  stellar  components  (e.g. major  `dry
mergers') of galaxies  cannot be a dominant mode of  stellar mass assembly for
galaxies of any stellar mass (on average).

\section{Summary}
\label{sec_discuss}

In  this paper  we have  presented model  predictions for  the  star formation
histories (SFHs)  of {\it central}  galaxies as a  function of halo  mass. The
model is  based on {\it  self-consistent} modeling of the  conditional stellar
mass function  across cosmic time by  Yang \etal (2012).   Two key ingredients
are used  in deriving the  SFHs. The first  is the mass assembly  histories of
central  galaxies and  their accreted  satellite galaxies,  and second  is the
local  observational  constraints  on  the  star formation  rates  of  central
galaxies  as function  of halo  mass.  The  difference in  total  stellar mass
between   the  accreted   and  surviving   satellites  provides   the  maximum
contribution available to  the growth of the central  galaxy through accretion
of stars  from satellites. The  minimum (zero) and maximum  contributions from
the accreted satellites correspond to the maximum and minimum amounts of stars
that  formed {\it  in  situ} in  a  central galaxy.   As  expected, the  local
observational  constraints  on the  star  formation  rates  (SFRs) of  central
galaxy, obtained from the SDSS DR7 group catalog, fall between these extrema.

Using  these  data, we  have  obtained median  SFHs  for  central galaxies  as
function  of their  present-day  halo  mass.  We  have  presented a  universal
fitting formula that adequately describes the dependence of these SFHs on halo
mass,  galaxy stellar  mass and  redshift.  We also  used this  model to  make
predictions  for (i)  the star  formation  rates (SFRs),  star formation  rate
densities  (SFRDs),  gas mass  depletion  rates  (GMDRs),  and star  formation
efficiencies (SFEs), all as functions of redshift, halo mass and stellar mass;
(ii) the cosmic  star formation rate density; and (iii)  the fraction of stars
that  have formed {\it  in situ}  over cosmic  time (as  apposed to  have been
accreted).  Our main findings can be summarized as follows.  

\begin{enumerate}

\item  The  SSFR  at  high  $z$ increases  rapidly  with  increasing  redshift
  [$\propto (1+z)^{2.5}$] for halos of a given mass, and slowly with halo mass
  ($\propto M_h^{0.12}$) for a given  $z$. This scaling is almost identical to
  that of the  specific mass accretion rate of dark  matter halos (Dekel \etal
  2009; McBride \etal 2009; Fakhouri \etal 2010), indicating that the 
  SFR of (star-forming) central galaxies is largely
  regulated by the rate at which their host halos accrete mass. Such a picture
  has strong theoretical  support (e.g., Dutton, van den  Bosch \& Dekel 2010;
  Bouch\'e \etal 2010; Dav\'e, Finlator \& Oppenheimer 2012).

\item The  ratio between the SFR in  the halo's main progenitor  and the final
  stellar mass of a galaxy peaks  roughly at a constant value, $\sim 10^{-9.3}
  h^2 {\rm  yr}^{-1}$, independent of  the halo mass  and stellar mass  of the
  galaxy at  the present day.  The  redshift at which this  SFR peaks ($z_{\rm
    pk}$), however,  increases rapidly with  the present-day halo mass  of the
  galaxy,  with $z_{\rm  pk}\sim 0.5$  for $M_h  =10^{11}\msunh$,  and $z_{\rm
    pk}\sim 3$ for $M_h =10^{15}\msunh$.

\item More than  half of the stars in the present-day  Universe were formed in
  halos  with masses  between $10^{11.1}\msunh$  and $10^{12.3}\msunh$  in the
  redshift range  $0.4$ - $1.9$.  Halos with  masses between $10^{11.5}\msunh$
  and  $10^{12.5}\msunh$  dominate the  star  formation  rate  density of  the
  universe over a large range of redshift, from $z\sim 1$ to $\sim5$; at $z<1$
  the star formation rate density is dominated by halos with $10^{10.5}\msunh<
  M_h <10^{11.5}\msunh$; the total amounts of stars formed in small halos with
  $M_h<10^{10.5}\msunh$  and in massive  halos with  $M_h>10^{12.5}\msunh$ are
  both negligibly small at any $z<5$.

\item For  individual centrals,  the SFE, defined  as the star  formation rate
  divided  by  the baryonic  accretion  rate,  initially  increases, until  it
  reaches a maximum, after which it decreases rapidly to a quenched state. For
  centrals  in present-day  halos with  $M_h \gta  10^{13}  \msunh$, quenching
  occurs  when their  main  progenitor  halo reaches  a  mass $\sim  10^{12.5}
  \msunh$, at which  point the SFE is $\sim 0.1$.  For centrals in present-day
  halos with $M_h \lta 10^{13}\msunh$  the quenching mass shifts to lower halo
  masses at  higher peak SFE;  at the present,  the quenching mass is  $\sim 2
  \times 10^{11} \msunh$, with a peak SFE close to unity.

\item Whereas the SFE histories of central galaxies that end-up in present-day
  halos with $M_h \lta 10^{12.5} \msunh$ are consistent with the presence of a
  halo  mass floor  of  $\sim 10^{11}  \msunh$,  as suggested  by Bouch\'e  et
  al. (2010),  our model indicates  that such a  halo mass floor  (below which
  star  formation is  suppressed) needs  to  be substantially  lower, or  even
  absent, at high $z$.

\item There are some indications  that our model may underpredict the cosmic
  star formation density  at high redshifts ($z \ga  3$).  The discrepancy can
  be significantly reduced if either the  stellar mass functions  at high
  redshift have  a significantly  steeper low-mass end slope, or the 
  faint-end slopes of the luminosity functions used in the SFD
  measurements are significantly under-estimated.

\item At redshift $z\ga 2.5$ more  than $99\%$ of the stars in the progenitors
  of massive galaxies (mainly ellipticals)  are formed {\it in situ}, and this
  fraction decreases  as a  function of redshift,  dropping to $\sim  60\%$ at
  $z=0$;  for a  Milky Way  sized halo  of $M_h\sim  10^{12}\msunh$  more than
  $80\%$  of  all  the  stars  in  the  central  galaxy  are  formed  {\it  in
    situ}.  Hence, major mergers  cannot be  a dominant  mode of  stellar mass
  assembly for any stellar mass (on average).

\end{enumerate}

\section*{Acknowledgements} 

We thank Shiyin Shen, Chunyan Jiang for help in dealing with IDL plotting, and
Stephane  Charlot for  kindly providing  us the  passive evolution  factors in
electronic  forms.  We  also  thank  the anonymous  referee  for  helpful
  comments that improved the presentation of this paper. This work is
supported  by  the  grants  from  NSFC (Nos.   10925314,  11128306,  11121062,
11233005)  and  CAS/SAFEA   International  Partnership  Program  for  Creative
Research Teams (KJCX2-YW-T23).   HJM would like to acknowledge  the support of
NSF AST-1109354 and NSF AST-0908334.


\appendix

\section{A. The SFH fitting formula for SMF2}
\begin{figure*}
\plotone{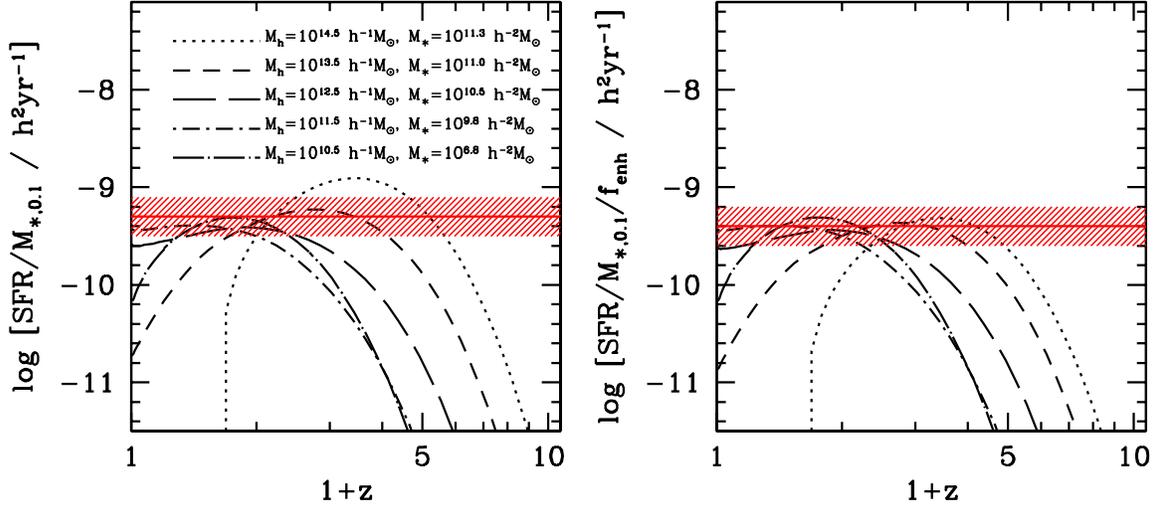}
\caption{Similar to  right-hand panel of  Fig.~\ref{fig:SFH_peak}, except that
  here the star formation histories are obtained using SMF2, rather than SMF1.
  In the left-hand panel the SFHs  are normalized by the stellar masses of the
  central galaxies at  $z=0.1$, while the curves in  the right-hand panel have
  been normalized by an additional enhancement factor, $f_{\rm enh}$, given by
  Eq.~(\ref{equ:SFR_fcor}).}
\label{fig:SFH_peak2}
\end{figure*}
\begin{figure*}
\plotone{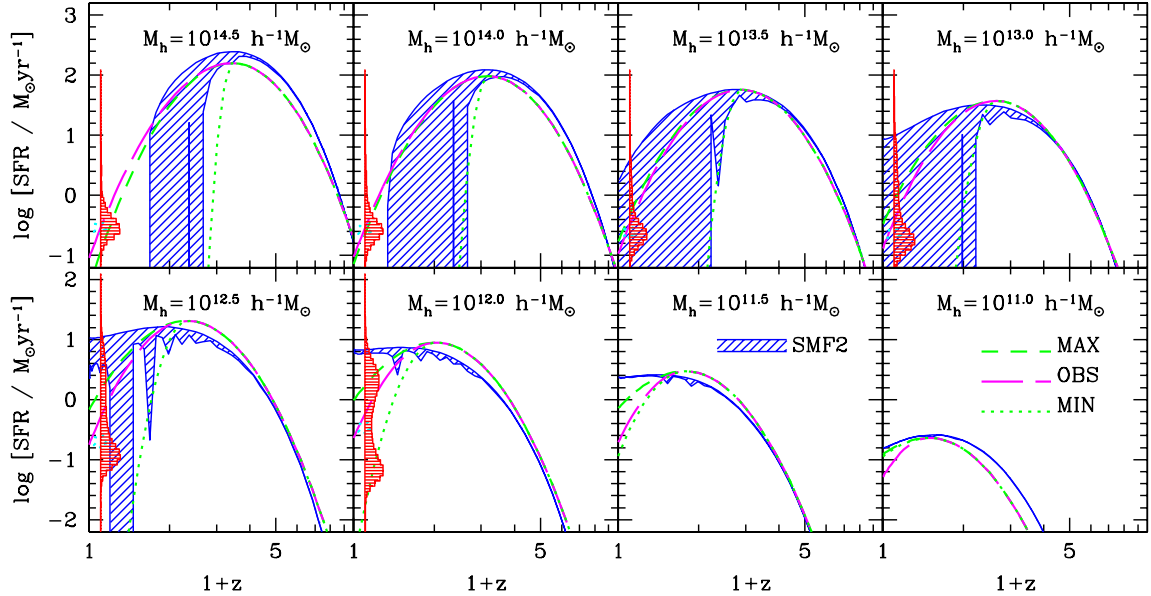}
\caption{The star formation  histories (SFHs) of central galaxies  in halos of
  different present-day masses, as indicated  in each panel. Here we only show
  the results obtained from SMF2.  In each panel, predictions with MIN and MAX
  assumptions for the  star formation rate are bridged  with shaded areas (see
  text for more details).  In panels for halos with $M_h \geq 10^{12} \msunh$,
  local observational constraints are  shown as the vertical shaded histograms
  (distributions)  and stars  (median).  The dotted,  dashed, and  long-dashed
  lines are the MIN, MAX and OBS fits to the SFHs discussed in the text.}
\label{fig:SFH_2}
\end{figure*}

The discussion in the main text is mainly based on SMF1.  As we have seen, the
SFHs  obtained from  SMF1 and  SMF2  have systematic  differences (see,  e.g.,
Fig.~\ref{fig:SFH}). For  completeness, this  Appendix presents our  model for
the SFHs  of central galaxies  based on SMF2,  rather than SMF1. Let  us first
look  at   the  amplitude   of  the   SFHs,  shown  in   the  left   panel  of
Fig.~\ref{fig:SFH_peak2}.  Compared to the results obtained from SMF1 (see the
right panel of Fig.~\ref{fig:SFH_peak}), the  peak values of the SFRs obtained
from  SMF2 have  slightly  larger variation  between  different halos  masses.
Especially the  very massive  galaxies in SMF2  have significantly  higher SFH
peaks than  those in SMF1.  In  order to properly  model the SFH peaks  in the
SMF2, we introduce a stellar  mass dependent enhancement factor $f_{\rm enh}$,
so that  we can  properly model the  peak amplitudes.   By fitting to  the SFH
peaks  obtained  from  SMF2,  we  get the  following  relation  between  ${\rm
  SFR}_{\rm pk}$ and $M_{\ast, 0.1}$: 
\begin{equation}\label{equ:SFR_amp2}
{{\rm SFR}_{\rm pk}\over [\msun\,{\rm yr}^{-1}] }
= {M_{\ast, 0.1} \over 10^{9.4}\msunhh} f_{\rm enh}(M_{\ast, 0.1})\,,
\end{equation}
where
\begin{equation}\label{equ:SFR_fcor}
 f_{\rm enh}(M_{\ast, 0.1}) = 1+(M_{\ast, 0.1}/10^{11.2}\msunhh)^2\,,
\end{equation}
is  a stellar  mass  dependent  enhancement factor.   The  performance of  the
fitting results is  shown in the right panel  of Fig. \ref{fig:SFH_peak2}.  As
one can see, the model describes the amplitudes remarkably well.

The shape of  the SFH obtained from  SMF2 can still be modeled  using the same
form as Eq.~(\ref{equ:SFH}), 
\begin{equation}\label{equ:SFH2}
{\rm SFR}(M_h, z) = {\rm SFR}_{\rm pk} 
\times\exp\left\{-{\log^2[(1+z)/(1+z_{\rm pk})]
\over 2\sigma^2(z_{\rm pk})}\right\}\,,
\end{equation}
and here with $\sigma(z_{\rm pk})$ given as follows.  For $z\ge z_{\rm pk}$,
\begin{equation}\label{equ:decay1_2}
\sigma (z_{\rm pk})= 0.146 (1+z_{\rm pk})^{-0.137}\,;
\end{equation}
while for $z< z_{\rm pk}$ it is
\begin{eqnarray}\label{equ:decay_2}
\sigma (z_{\rm pk})= \left\{\begin{array}{ll} 
 0.0857 (1+z_{\rm pk})^{0.391}~~~ &\mbox{(OBS)}\\
 0.168 (1+z_{\rm pk})^{-0.208}~~~ &\mbox{(MAX)} \\
 0.346 (1+z_{\rm pk})^{-2.18}~~~ &\mbox{(MIN)}
\end{array}\right.\,.
\end{eqnarray}
The  predictions of this  fitting model  are shown  as the  long-dashed (OBS),
dashed (MAX)  and dotted  (MIN) curves in  Fig. \ref{fig:SFH_2}  in comparison
with the results obtained directly from SMF2.

\end{document}